\documentclass[iop,apj,twocolappendix]{emulateapj}

\usepackage{natbib}
\usepackage{amsmath,amssymb}
\usepackage{apjfonts}
\usepackage{xspace}
\usepackage{rotating}
\usepackage{multirow}
\usepackage{dcolumn}

\usepackage[usenames]{xcolor}
\definecolor{mpl_blue}{HTML}{1F77B4}
\definecolor{mpl_orange}{HTML}{FF7F0E}
\definecolor{mpl_green}{HTML}{2CA02C}
\definecolor{mpl_red}{HTML}{D62728}

\newcommand{\num}[2]{#1\times 10^{#2}} 

\usepackage[backref, breaklinks, plainpages=false, colorlinks=true, anchorcolor=cyan, linkcolor=mpl_red, citecolor=cyan, urlcolor=magenta, bookmarks=false]{hyperref}
\usepackage[all]{hypcap}
\usepackage[caption=false]{subfig}

\shorttitle{NANOGrav $11$-year Gravitational-wave Background Constraints}
\shortauthors{The NANOGrav Collaboration}

\newcommand{\bayesephem}{\textsc{BayesEphem}}

\def\be{\begin{equation}}
\def\ee{\end{equation}}
\newcommand{\bb}{\begin{bmatrix}}
\newcommand{\eb}{\end{bmatrix}}

\def\bea{\begin{eqnarray}}
\def\eea{\end{eqnarray}}

\DeclareMathOperator{\Tr}{Tr}

\defcitealias{mop14}{MOP14}
\defcitealias{rws+14}{RWS14}
\defcitealias{s13}{S13}
\defcitealias{s16}{S16}
\defcitealias{dfg+13}{NG5}
\defcitealias{abb+15}{NG9a}
\defcitealias{abb+16}{NG9b}
\defcitealias{abb+17}{NG11}
\defcitealias{ltm+15}{LTM15}
\defcitealias{kh13}{KH13}
\defcitealias{mm13}{MM13}

\begin{document}

\title{The NANOGrav $11$-year Data Set: \\ Pulsar-Timing Constraints On The Stochastic Gravitational-Wave Background}

\author{
Z.~Arzoumanian\altaffilmark{1},
P.~T.~Baker\altaffilmark{2,3},
A.~Brazier\altaffilmark{4},
S.~Burke-Spolaor\altaffilmark{2,3},
S.~J.~Chamberlin\altaffilmark{5},
S.~Chatterjee\altaffilmark{4},
B.~Christy\altaffilmark{6},
J.~M.~Cordes\altaffilmark{4},
N.~J.~Cornish\altaffilmark{7},
F.~Crawford\altaffilmark{8},
H.~Thankful~Cromartie\altaffilmark{9},
K.~Crowter\altaffilmark{10},
M.~DeCesar\altaffilmark{$\dagger$11},
P.~B.~Demorest\altaffilmark{12},
T.~Dolch\altaffilmark{13},
J.~A.~Ellis\altaffilmark{$\dagger$2,3,$\ast\ast$}
R.~D.~Ferdman\altaffilmark{14},
E.~Ferrara\altaffilmark{15},
W.~M.~Folkner\altaffilmark{16},
E.~Fonseca\altaffilmark{17},
N.~Garver-Daniels\altaffilmark{2,3},
P.~A.~Gentile\altaffilmark{2,3},
R.~Haas\altaffilmark{18},
J.~S.~Hazboun\altaffilmark{$\dagger$19,20},
E.~A.~Huerta\altaffilmark{18,21},
K.~Islo\altaffilmark{22},
G.~Jones\altaffilmark{23},
M.~L.~Jones\altaffilmark{2,3},
D.~L.~Kaplan\altaffilmark{22},
V.~M.~Kaspi\altaffilmark{17},
M.~T.~Lam\altaffilmark{$\dagger$2,3},
T.~J.~W.~Lazio\altaffilmark{16,24},
L.~Levin\altaffilmark{25},
A.~N.~Lommen\altaffilmark{26},
D.~R.~Lorimer\altaffilmark{2,3},
J.~Luo\altaffilmark{20},
R.~S.~Lynch\altaffilmark{27},
D.~R.~Madison\altaffilmark{$\ddagger$28},
M.~A.~McLaughlin\altaffilmark{2,3},
S.~T.~McWilliams\altaffilmark{2,3},
C.~M.~F.~Mingarelli\altaffilmark{29},
C.~Ng\altaffilmark{10},
D.~J.~Nice\altaffilmark{11},
R.~S.~Park\altaffilmark{16},
T.~T.~Pennucci\altaffilmark{$\dagger$2,3,30,31},
N.~S.~Pol\altaffilmark{2,3},
S.~M.~Ransom\altaffilmark{9,28},
P.~S.~Ray\altaffilmark{32},
A.~Rasskazov\altaffilmark{30,31,33},
X.~Siemens\altaffilmark{22},
J.~Simon\altaffilmark{16},
R.~Spiewak\altaffilmark{22,34},
I.~H.~Stairs\altaffilmark{10},
D.~R.~Stinebring\altaffilmark{35},
K.~Stovall\altaffilmark{$\dagger$12},
J.~Swiggum\altaffilmark{$\dagger$22},
S.~R.~Taylor\altaffilmark{${\color{magenta}\S}\dagger$24,16},
M.~Vallisneri\altaffilmark{16,24},
R.~van Haasteren\altaffilmark{16$\ast$},
S.~Vigeland\altaffilmark{$\dagger$22},
W.~W.~Zhu\altaffilmark{36,37}\\
(The NANOGrav Collaboration)\altaffilmark{$\star$}}

\affil{$\star$Author order alphabetical by surname}
\affil{$^{1}$Center for Research and Exploration in Space Science and Technology and X-Ray Astrophysics Laboratory,\\ NASA Goddard Space Flight Center, Code 662, Greenbelt, MD 20771, USA}
\affil{$^{2}$Department of Physics and Astronomy, West Virginia University, P.O.~Box 6315, Morgantown, WV 26506, USA}
\affil{$^{3}$Center for Gravitational Waves and Cosmology, West Virginia University, Chestnut Ridge Research Building, Morgantown, WV 26505, USA}
\affil{$^{4}$Department of Astronomy, Cornell University, Ithaca, NY 14853, USA}
\affil{$^{5}$Department of Astronomy and Astrophysics, Pennsylvania State University, University Park, PA 16802, USA}
\affil{$^{6}$Notre Dame of Maryland University, 4701 N.~Charles Street, Baltimore, MD 21210, USA}
\affil{$^{7}$Department of Physics, Montana State University, Bozeman, MT 59717, USA}
\affil{$^{8}$Department of Physics and Astronomy, Franklin \& Marshall College, P.O.~Box 3003, Lancaster, PA 17604, USA}
\affil{$^{9}$University of Virginia, Department of Astronomy, P.O.~Box 400325, Charlottesville, VA 22904, USA}
\affil{$^{10}$Department of Physics and Astronomy, University of British Columbia, 6224 Agricultural Road, Vancouver, BC V6T 1Z1, Canada}
\affil{$^{11}$Department of Physics, Lafayette College, Easton, PA 18042, USA}
\affil{$^{12}$National Radio Astronomy Observatory, 1003 Lopezville Rd., Socorro, NM 87801, USA}
\affil{$^{13}$Department of Physics, Hillsdale College, 33 E.~College Street, Hillsdale, Michigan 49242, USA}
\affil{$^{14}$Department of Physics, University of East Anglia, Norwich, UK}
\affil{$^{15}$NASA Goddard Space Flight Center, Greenbelt, MD 20771, USA}
\affil{$^{16}$Jet Propulsion Laboratory, California Institute of Technology, 4800 Oak Grove Drive, Pasadena, CA 91109, USA}
\affil{$^{17}$Department of Physics, McGill University, 3600  University St., Montreal, QC H3A 2T8, Canada}
\affil{$^{18}$NCSA, University of Illinois at Urbana-Champaign, Urbana, Illinois 61801, USA}
\affil{$^{19}$University of Washington Bothell, 18115 Campus Way NE, Bothell, WA 98011, USA}
\affil{$^{20}$Center for Advanced Radio Astronomy, University of Texas Rio Grande Valley, Brownsville, TX 78520, USA}
\affil{$^{21}$Department of Astronomy, University of Illinois at Urbana-Champaign, Urbana, Illinois 61801, USA}
\affil{$^{22}$Center for Gravitation, Cosmology and Astrophysics, Department of Physics, University of Wisconsin-Milwaukee,\\ P.O.~Box 413, Milwaukee, WI 53201, USA}
\affil{$^{23}$Department of Physics, Columbia University, New York, NY 10027, USA}
\affil{$^{24}$Theoretical AstroPhysics Including Relativity (TAPIR), MC 350-17, California Institute of Technology, Pasadena, California 91125, USA}
\affil{$^{25}$Jodrell Bank Centre for Astrophysics, University of Manchester, Manchester, M13 9PL, United Kingdom}
\affil{$^{26}$Department of Physics and Astronomy, Haverford College, Haverford, PA 19041, USA}
\affil{$^{27}$Green Bank Observatory, P.O.~Box 2, Green Bank, WV 24944, USA}
\affil{$^{28}$National Radio Astronomy Observatory, 520 Edgemont Road, Charlottesville, VA 22903, USA}
\affil{$^{29}$Center for Computational Astrophysics, Flatiron Institute, 162 Fifth Avenue, New York, NY 10010, USA}
\affil{$^{30}$Institute of Physics, E\"{o}tv\"{o}s Lor\'{a}nd University, P\'{a}zm\'{a}ny P. s. 1/A, 1117 Budapest, Hungary}
\affil{$^{31}$Hungarian Academy of Sciences MTA-ELTE Extragalactic Astrophysics Research Group, 1117 Budapest, Hungary}
\affil{$^{32}$Naval Research Laboratory, Washington DC 20375, USA}
\affil{$^{33}$School of Physics and Astronomy and Center for Computational Relativity and Gravitation,\\ Rochester Institute of Technology, Rochester, NY 14623, USA}
\affil{$^{34}$Centre for Astrophysics and Supercomputing, Swinburne University of Technology, PO Box 218, Hawthorn VIC 3122, Australia}
\affil{$^{35}$Department of Physics and Astronomy, Oberlin College, Oberlin, OH 44074, USA}
\affil{$^{36}$National Astronomical Observatories, Chinese Academy of Science, 20A Datun Road, Chaoyang District, Beijing 100012, China}
\affil{$^{37}$Max Planck Institute for Radio Astronomy, Auf dem H\"{u}gel 69, D-53121 Bonn, Germany}
\affil{$^{\dagger}$ NANOGrav Physics Frontiers Center Postdoctoral Fellow}
\affil{$^{\ddagger}$ Jansky Fellow}
\affil{$\ast$ Currently employed at Microsoft Corporation}
\affil{$\ast\ast$ Currently employed at Infinia ML, 202 Rigsbee Avenue, Durham NC, 27701}
\email[${\color{magenta}\S}$ Corresponding author: S.~R.~Taylor, ]{srtaylor@caltech.edu}

\begin{abstract}
We search for an isotropic stochastic gravitational-wave background (GWB) in the newly released $11$-year dataset from the North American Nanohertz Observatory for Gravitational Waves (NANOGrav). While we find no evidence for a GWB, we place constraints on a population of inspiraling supermassive black hole (SMBH) binaries, a network of decaying cosmic strings, and a primordial GWB. For the first time, we find that the GWB constraints are sensitive to the Solar System ephemeris (SSE) model used, and that SSE errors can mimic a GWB signal. We developed an approach that bridges systematic SSE differences, producing the first PTA constraints that are robust against SSE errors. We thus place a $95\%$ upper limit on the GW strain amplitude of $A_\mathrm{GWB}<1.45\times10^{-15}$ at a frequency of $f=1$-yr$^{-1}$ for a fiducial $f^{-2/3}$ power-law spectrum, and with inter-pulsar correlations modeled. This is a factor of $\sim2$ improvement over the NANOGrav $9$-year limit, calculated using the same procedure. Previous PTA upper limits on the GWB (as well as their astrophysical and cosmological interpretations) will need revision in light of SSE systematic errors. We use our constraints to characterize the combined influence on the GWB of the stellar mass-density in galactic cores, the eccentricity of SMBH binaries, and SMBH--galactic-bulge scaling relationships. We constrain cosmic-string tension using recent simulations, yielding an SSE-marginalized $95\%$ upper limit of $G\mu<5.3\times10^{-11}$---a factor of $\sim2$ better than the published NANOGrav $9$-year constraints. Our SSE-marginalized $95\%$ upper limit on the energy density of a primordial GWB (for a radiation-dominated post-inflation Universe) is $\Omega_\mathrm{GWB}(f)h^2<3.4\times10^{-10}$.
\end{abstract}

\section{Introduction}
\label{sec:intro}

Over the last two years, the gravitational-wave (GW) community celebrated the first direct detection of GWs, generated by the coalescence of two $\sim 30 \, \mathrm{M}_\odot$ black holes \citep{aaa+16}, as well as the first multi-messenger observation of GWs with pan-spectral EM radiation, emitted during and after the final inspiral and merger of two neutron stars \citep{aaa+17}.
Pulsar-timing arrays [PTAs, \citep{saz78, det79, fb90}] offer the opportunity of extending GW observations to the very-low-frequency spectrum ($\sim 1$--$100$ nHz). The discovery-space here is populated by GWs from supermassive black-hole binaries (SMBHBs) at galactic centers \citep[see e.g.][]{shm+04,s13c}, and possibly from more speculative sources of cosmological origin, such as cosmic strings \citep{dv01,oms10} and/or a primordial GW background (GWB) produced by quantum fluctuations of the gravitational field in the early Universe, amplified by inflation, e.g. \cite{g05, lms+16}.
 
The three major collaborations involved in this effort are the North American Nanohertz Observatory for Gravitational-waves (NANOGrav, \citet{ml13}), the European Pulsar Timing Array (EPTA, \citet{dcl+16}), and the Parkes Pulsar Timing Array (PPTA, \citet{h13}). Additionally, the International Pulsar Timing Array (IPTA, \citet{v+16}) exists as an umbrella consortium for data-sharing, coordinated timing campaigns, and joint GW analysis. The increasing sensitivity of PTAs is apparent in the ever-tightening upper limits \citep{vhj+11,dfg+13,src+13,ltm+15,srl+15,abb+16} on the stochastic GWB from the unresolved superposition of SMBHB signals out to redshift $\lesssim 1$.

The road toward detection lies not only through the accumulation of ever longer and more accurate time-of-arrival (TOA) data for larger arrays of monitored pulsars, but also through the development of powerful, robust, and reliable data-analysis methods to demonstrate the presence of GWs in PTA data.
In this article, we report substantial advances along both avenues.
First, we present our stochastic-GW analysis of NANOGrav's largest and most sensitive dataset so far, spanning $45$ pulsars and $11.4$ years.
See Sec.\ \ref{sec:obs} and \citet[][hereafter \citetalias{abb+17}]{abb+17} for more about this ``11-year'' dataset. Second, we describe our statistical-inference framework, which was significantly augmented compared to our GW study of the 9-year dataset \citep[][hereafter \citetalias{abb+16}]{abb+16}.
Improvements include a practical strategy to isolate the expected signature of stochastic GWs in our data---namely the emergence of a long-timescale noise process that is common to all pulsars, and the positive detection of inter-pulsar spatial correlations with a quadrupolar signature \citep{hd83}. This strategy is based on Bayesian model selection, and is extensible to large arrays and datasets. Indeed, for the first time with a large pulsar array, we are able to report GW upper limits and GW-vs-noise (``detection'') Bayes factors computed with likelihoods that include spatial correlations -- such as the ones predicted by \cite{hd83} -- a goal that had previously proved computationally unfeasible beyond small arrays \citep{ltm+15}.

This article also features a more robust, Bayesian--frequentist hybrid ``optimal-statistic'' analysis \citep{abc+09,dfg+13,ccs+15}, which complements our primary Bayesian approach. Additionally, we employ a more flexible end-to-end approach for PTA GW searches to constrain astrophysical parameters (characterizing SMBHB populations and environments, as well as cosmic-string properties). This approach uses a set of GW-spectrum simulations that span the parameter-space region of interest, and interpolates them by means of Gaussian processes (GPs) \citep{rw06, tss17}, resulting in a flexible new model that is calibrated directly by detailed simulations. 

Lastly, but perhaps most importantly, we report on how Solar System ephemeris (SSE) errors can manifest as a false GWB signal in PTA data, for sufficiently long and high-quality datasets. The SSE is used to refer TOA measurements to an inertial frame located at the Solar System barycenter (SSB). 
Previous GW searches treated the ephemeris as a fixed-parameter model without uncertainties. However, in the course of analyzing the 11-year dataset we discovered that adopting different ephemerides (among the last few published by the Jet Propulsion Laboratory (JPL); see \citealt{2009IPNPR.178C...1F,2014IPNPR.196C...1F,de435,de436}) leads to significantly different upper-limit and model-comparison statistics.
As PTA datasets become larger, longer, and more precise, our GW searches will continue to uncover systematic effects that will limit our sensitivity unless handled appropriately.
To this end, we have developed a physical model of ephemeris uncertainties, and we demonstrate that it makes our analysis insensitive to the choice among recent ephemerides.

This paper is laid out as follows:  methodological advances are discussed in Sec.\ \ref{sec:data_analysis}.
In Secs.\ \ref{sec:results} and \ref{sec:discussion} we report GW upper limits and detection Bayes factors based on the 11-year dataset, as well as new constraints on astrophysical and cosmological sources of low-frequency GWs. In Sec.\ \ref{sec:conclusions} we present our conclusions and discuss prospects for future observations.

For the busy reader, the following summarizes the most consequential results:
\begin{itemize}
\item Once we take ephemeris uncertainty into account, we find Bayesian model comparison to be inconclusive on the presence of a GWB-like signal in the data (with signal-vs.-noise and spatial-correlation Bayes factors both $\sim 1$). Adopting one of the fixed JPL ephemerides leads to signal-vs.-noise Bayes factors as high as $26 \pm 2$ in favor of a GWB-like signal (for JPL ephemeris DE430), suggesting that systematic ephemeris errors can masquerade as GWs---and conversely that modeling these errors can subtract power from a putative GWB signal.
This degeneracy will be resolved over the next few years as we collect longer and larger datasets, and as ephemeris accuracy improves with data from current NASA missions.
\item Accounting for ephemeris uncertainty, the 95\% Bayesian upper limit on a fiducial $f^{-2/3}$ GW spectrum from SMBHBs is $A_\mathrm{GW}^{95\%} = 1.45(2) \times 10^{-15}$ at $f = 1/\mathrm{yr}$ (when modeling spatial correlations; $1.34(1) \times 10^{-15}$ when omitting them).
This value is modestly improved from the 9-year result of $A_\mathrm{GW}^{95\%} = 1.5 \times 10^{-15}$, which omitted correlations, and assumed JPL ephemeris DE421 as a fixed-parameter model without uncertainties.
Note however that reprocessing the 9-year dataset accounting for ephemeris uncertainties leads to $A_\mathrm{GW}^{95\%} = 2.91(2) \times 10^{-15}$ (when modeling spatial correlations; $2.67(2) \times 10^{-15}$ omitting them).
We expect that recently published limits from other PTAs (such as \citealt{srl+15}) would be likewise revised upwards.
Our 11-year upper limits assuming individual fixed-model ephemerides range from 1.53 to $1.78 \times 10^{-15}$ (when modeling spatial correlations), again suggesting that ephemeris errors can mimic GW-like signals.
\item We place the first joint constraints on the galaxy properties and binary evolution parameters with the greatest impact on the spectral shape and amplitude of the GWB from SMBHBs. Previous work, such as that undertaken in \citetalias{abb+16}, has always utilized an amplitude or spectral shape assumption before inferring any astrophysical constraints. This improved methodology allows for the first robust PTA limits on the $M_\mathrm{BH} - M_\mathrm{bulge}$ relation, and shows that the NANOGrav $11$-year dataset prefers a relation that is lower (in terms of the relation's $y$-intercept) than that reported in \citet{kh13}.
\item Using a model of cosmic-string--generated GW spectra that interpolates among extensive string-network simulations \citep{bo17}, we place a 95\% upper limit of $5.3(2) \times 10^{-11}$ on the string tension $G\mu/c^2$ for a reconnection probability $p = 1$. This result is marginalized over ephemeris uncertainties, but neglects inter-pulsar spatial correlations.
(Including these is still too taxing computationally; however we argue that our upper limits assuming a variable power-law exponent, described in Sec.\ \ref{sec:balimit}, are affected modestly by correlations, and so should be the cosmic-string result.) Previous studies reported limits of $1.3 \times 10^{-10}$ \citepalias{abb+16} and $8.6 \times 10^{-10}$ \citep{ltm+15}, although different prior assumptions and the lack of ephemeris modeling preclude a direct comparison.
\item Lastly, we can interpret power-law GWBs with different fixed exponents as a primordial background amplified through inflation, with post-inflationary eras characterized by different equations of state.
Assuming a radiation-dominated post-inflation Universe, and a tensor index $n_t = 0$ (corresponding to a scale-invariant spectrum) leads to a 95\% upper limit of $3.4(1) \times 10^{-10}$ on the GW energy density $\Omega_\mathrm{GW}(f)h^2$ at $f = 1\,\mathrm{yr}^{-1}$, again marginalizing over ephemeris uncertainty but neglecting inter-pulsar spatial correlations.
\end{itemize}

\section{The $11$-year Data Set}
\label{sec:obs}

%
\begin{figure}
  \includegraphics[width=\columnwidth]{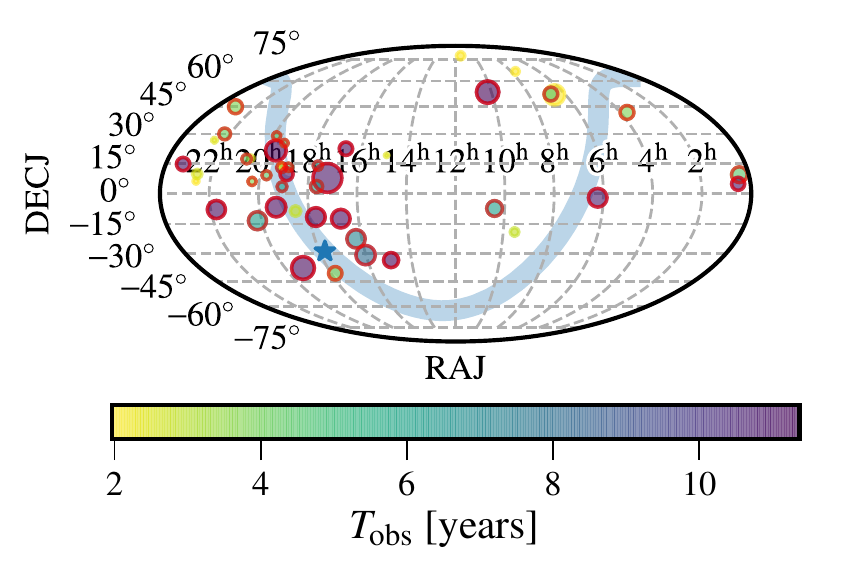} \\
  \caption{Sky positions of all $45$ pulsars in the NANOGrav $11$-year dataset. The area of each circle is indicative of the number of TOAs, while the color scale indicates the observational baseline. The $34$ pulsars whose baselines are longer than three years are indicated with solid red edges. The Milky Way plane is shown behind as a blue band (thickness is not indicative of Galactic scale height), with the Galactic center shown as a blue star. The longest baseline is given by J$1744$$-$$1134$ with $11.37$ years, while the largest dataset is given by J$1713$$+$$0747$ with 27571 TOAs.}
  \label{fig:pulsar_dist}
\end{figure}

Our analyses throughout this paper make use of the NANOGrav $11$-year dataset, which consists of the TOA data and pulsar timing models recently presented in \citetalias{abb+17}, and is publicly available online\footnote{\href{http://data.nanograv.org}{data.nanograv.org}}.
This dataset is derived from timing observations of $45$ millisecond pulsars between July 30th, 2004 to December 31st, 2015.
The first five years of data on seventeen pulsars constituted the NANOGrav $5$-year dataset, which we previously published in \citetalias{dfg+13}.
The $5$-year dataset was augmented with four years of data, reported as the $9$-year dataset in \citet[][hereafter \citetalias{abb+15}]{abb+15}, which came with the substantial improvements of new broadband instrumentation, a nearly twofold increase in the timing baseline for the original $17$ pulsars, and a more than twofold increase in the total number of observed sources to $37$ pulsars.
The present extension of the $9$-year dataset is composed of two years of data that were observed and processed in a nearly identical fashion to the previous augmentation, with the addition of nine pulsars and the removal of one (see \citetalias{abb+17} for full details). 
Here we briefly review the instrumentation, observations, and basic data reduction of the entire dataset, referring the reader to \citetalias{abb+17}, \citetalias{abb+15}, and references therein for a thorough description. A sky map of of the pulsars in this data set is shown in \autoref{fig:pulsar_dist}, with indicators of the time span and data volume for each pulsar. 

\subsection{Observations}

We obtained all data using the 100-m Robert C. Byrd Green Bank Telescope (GBT) of the Green Bank Observatory\footnote{\href{http://greenbankobservatory.org/telescopes/gbt/}{greenbankobservatory.org/telescopes/gbt/}} and the 305-m William E. Gordon Telescope (Arecibo) of Arecibo Observatory\footnote{\href{http://outreach.naic.edu/ao/}{outreach.naic.edu/ao/}}.
Sources within Arecibo's declination range ($0^\circ < \delta < 39^\circ$) were observed there due to its superior sensitivity, and only pulsars J1713+0747 and B1937+21 were observed at both telescopes.
Excluding early portions from the $5$-year dataset, we observed each source roughly once a month for the entire dataset.
In addition, some pulsars have been observed weekly in a campaign to increase our sensitivity to individual sources of GWs \citep[][Section 6.1]{abb+14}. Specifically, two pulsars have been observed weekly at the GBT since $2013$ (PSRs J1737$+$0747 and J1909$-$3744), and five pulsars have been observed weekly at Arecibo since $2015$ (PSRs J0030$+$0451, J1640$+$2224, J1713$+$0747, J2043$+$1711, and J2317$+$1439).

During most epochs,\footnote{This excludes the weekly observations, which were performed at 1.4 GHz only, as well as epochs for which receivers were unavailable for technical reasons.} we observed sources in two widely separated frequency bands, in order to accurately remove the frequency-dependent dispersion delay introduced by the ionized interstellar medium (ISM).
At the GBT, we used the 820-MHz and 1.4-GHz receivers for all observations. Since mechanical and time constraints prohibit alternating continually between the two receivers, observations in the two bands were always separated, typically by several days.
At Arecibo, we observed all pulsars at 1.4 GHz, plus a second frequency band (centered at either 430 MHz or 2.3 GHz) chosen depending on the spectrum and ISM characteristics of each pulsar\footnote{Pulsar J2317+1439 was originally observed with the 327 and 430-MHz receivers, but in 2014 we replaced the former with the 1.4-GHz receiver.}.
Pulsars observed at Arecibo are always observed in the two frequency bands one after another, separated by a few minutes.

For approximately the first six years, data were acquired with an identical pair of backend instruments that have since been decommissioned (GASP at the GBT, ASP at Arecibo).
Since 2010 and 2012, respectively, the broadband-capable backend clones GUPPI (at the GBT) and PUPPI (at Arecibo) have been used for taking data.

\subsection{Processing \& Time-Of-Arrival Data}

The raw data products are folded light curves (average, uncalibrated flux density as a function of rotational phase, divided into 2,048 phase bins) as a function of time, radio frequency, and polarization.
These data were cleaned of radio-frequency interference in several steps, polarization calibrated according to standard techniques, and averaged in time and frequency.
The final time resolution was either 30 minutes or 2.5\% of the binary period, whichever was shorter (approximately two thirds of our pulsars are in binary systems). This length of time corresponds to how long we point our radio telescopes at each pulsar during a single observation session, where we fold many individual pulses then convolve with a pulse-profile template to compute a single TOA. This is necessary to achieve high timing precisions $\sim 100\,\mathrm{ns}$.
The final frequency resolution varied between 1.5 and 12.5~MHz, depending on the receiver--backend combination.

The $5$-year TOA dataset was left mostly untouched as a subset of the $11$-year dataset, except for reprocessing under DE$436$. All of the GUPPI and PUPPI profile data, however, were reprocessed from scratch to make a consistent set of TOAs.
The TOAs were generated using standard template-matching cross-correlation methods, using only the total intensity profiles, producing one TOA per frequency channel per temporal subintegration.
Existing template profiles were reused for pulsars that were part of the $9$-year dataset, and created for new pulsars.

An additional set of procedures culled ``outlier'', low signal-to-noise (non-Gaussian distributed), or otherwise corrupt TOAs from the dataset using methods described in \cite{vvh17}.
The $11$-year dataset comprises a total of $309,201$~TOAs. 
All data reduction was completed using PSRCHIVE\footnote{\href{http://psrchive.sourceforge.net}{psrchive.sourceforge.net}} \citep{hvsm04} and custom NANOGrav processing scripts\footnote{\href{http://github.com/demorest/nanopipe}{github.com/demorest/nanopipe}}.

\subsection{Timing Models \& Noise Analysis}

Timing models from the $9$-year dataset were refit to the extended dataset and updated to include new parameters when deemed necessary on the basis of statistical significance tests.
We fit timing models for newly-added pulsars using a procedure similar to that described in \citetalias{abb+15}.
All timing models were created or updated using the standard timing software TEMPO\footnote{\href{http://tempo.sourceforge.net}{tempo.sourceforge.net}} and TEMPO2\footnote{\href{https://bitbucket.org/psrsoft/tempo2.git}{bitbucket.org/psrsoft/tempo2.git}} \citep{hem06,ehm06}, and crosschecked for consistency.

A standard noise model was also fit simultaneously with the timing model as described in \citetalias{abb+15} and \citetalias{abb+17}.
Each pulsar's white noise model includes a scale parameter on the TOA uncertainties (EFAC), an added variance (EQUAD), and a per-epoch variance (ECORR) for each observing system (i.e.\ a unique combination of backend and receiver).
In addition, a red noise process for each pulsar was modeled by a power-law spectral density described by an amplitude and spectral index. The inclusion of a red process in the noise model was not favored by all pulsars, but we include it in all subsequent analyses since this does not affect parameter constraints.
In the analyses described in the subsequent sections, we vary the pulsars' red noise parameters and the parameters of the gravitational wave background, but fix the white noise parameters.
Allowing the white noise parameters to vary does not alter the results, but significantly increases the computation time.

The SSE model used for the original analysis of the $5$-year dataset \citep{dfg+13} was DE$405$ \citep{de405}, while for the $9$-year dataset \citep{abb+15} all data (whether new, or from the $5$-year dataset) was modeled with DE$421$ \citep{2009IPNPR.178C...1F}. For the $11$-year dataset we use DE$436$ \citep{de436} as the fiducial SSE under which the data is processed and released. We do not need separate dataset releases for the different SSEs that we investigate in the following, since our GWB analysis incorporates marginalization over all affected processes, such as the individual timing and red-noise models.

\section{Data Analysis Methods}
\label{sec:data_analysis}

Characterizing all deterministic and noise processes in each pulsar, as well as teasing out a putative GWB signature from the cross-correlation of large datasets, requires a robust and sophisticated statistical framework. In the following we describe the major new features of the NANOGrav PTA analysis framework, as updated from \citetalias{abb+16}. Sec.\ \ref{sec:bayes} describes our use of Bayesian inference as it pertains to computing GWB upper limits and detection statistics. Sec.\ \ref{sec:spectralmodels} outlines how a GWB manifests in our data as a long-timescale stochastic process with a distinctive correlation signature between pulsars. In Sec.\ \ref{sec:eph} we describe how the Solar System ephemeris model appears in our pulsar-timing analysis, and our new Bayesian scheme to mitigate its uncertainties. The structure of our generative signal and noise model is outlined in Sec.\ \ref{sec:ptalike}, followed in Sec.\ \ref{sec:optimalstatdef} by the definition of our frequentist estimator for the GWB amplitude and significance. Finally, in Sec.\ \ref{sec:software} we list and provide links for all open-source software used in our GWB analysis.

\subsection{Bayesian methods}
\label{sec:bayes}

We primarily employ Bayesian inference (see, e.g., \citealt{gregory05book}) to extract physical information from our data, deriving marginalized posterior distributions and credible regions, basing upper limits on credible intervals, and relying on ratios of evidences (a.k.a.\ Bayes factors) to compare models with different assumptions and parametrizations. We explore our high-dimensional parameter space stochastically, using the parallel-tempering Markov Chain Monte Carlo (MCMC) sampler \citep{evh17b} described in the appendices of \citet{abb+14}.

To place upper limits on the GWB amplitude $A_\mathrm{GWB}$, we compute its posterior density distribution $p(A_\mathrm{GWB}|\mathcal{D})$ (with $\mathcal{D}$ the data) by giving $A_\mathrm{GWB}$ a uniform prior distribution that encloses the support of the likelihood, and we estimate the 95\% quantile by means of the empirical cumulative-distribution function estimator $\hat{A}^{95\%}_\mathrm{GWB}$ \citep{w26}. We approximate the standard error of $\hat{A}^{95\%}_\mathrm{GWB}$ as
\begin{equation}
\frac{\sqrt{x (1-x) / N}}{p(A_\mathrm{GWB}=\hat{A}^{95\%}_\mathrm{GWB}|\mathcal{D})},
\end{equation}
with $x = 0.95$ and $N$ the number of (quasi-)independent samples\footnote{Quasi-independence here refers to samples separated by one auto-correlation chain length.} in the chain.

As our PTA dataset becomes longer and more sensitive, we expect that evidence for the presence of GWs will emerge in two phases: first, as red-spectrum processes with the same amplitude in each pulsar, and with spectral slope consistent with an SMBHB population; later (perhaps several years), and conclusively, as Hellings--Downs spatial correlations predicted for an isotropic GWB. We note that anisotropic GWBs will have different (but predictable) spatial correlations \citep{msmv13, tg13, ms14, grt+14}.

Correspondingly, we characterize evidence for a GWB in the 11-year dataset in two steps. We first obtain the Bayes factor for a dataset model that includes a red-spectrum process with common statistical properties in all pulsars (but is uncorrelated between them), against a model with only per-pulsar noise processes. This is signal-vs.-noise model selection. We then obtain the Bayes factor for Hellings--Downs inter-pulsar spatial correlations vs.\ no correlations at all. This is spatial-correlation model selection, which we consider the definitive scheme for GWB detection. We also perform variants of these comparisons---for instance, we compare the Hellings--Downs and uncorrelated process against processes with monopolar (akin to long-timescale clock errors) and dipolar (akin to SSE errors) spatial correlations.

In all cases, we adopt bounded log-uniform priors for $A_\mathrm{GWB}$ and all other red-process amplitudes.
We adopt two different techniques to compute Bayes factors, according to the relation between the models in the comparison.

For \emph{nested} models (in our case, a signal-plus-noise model $\mathcal{H}_1$ and a noise-only model $\mathcal{H}_0$ obtained by fixing the GW amplitude to 0) we employ the Savage--Dickey formula \citep{d71}
\begin{equation}
\mathcal{B}_{10} \equiv \frac{\text{evidence}[\mathcal{H}_1]}{\text{evidence}[\mathcal{H}_0]} = \frac{p(A_\mathrm{GWB} = 0|\mathcal{H}_1)}{p(A_\mathrm{GWB} = 0|\mathcal{D},\mathcal{H}_1)},
\label{eq:savagedickey}
\end{equation}
where the numerator and denominator are, respectively, the prior and posterior probability density of $A_\mathrm{GWB} = 0$ in the embedding model $\mathcal{H}_1$.
We generate a sampling of $p(A_\mathrm{GWB}|\mathcal{D},\mathcal{H}_1)$ via MCMC, and we approximate $p(A_\mathrm{GWB} = 0|\mathcal{D},\mathcal{H}_1)$ as the normalized fraction of samples in the lowest-amplitude bin, averaging the estimate over a range of bin sizes. The standard error of this average yields an estimate of uncertainty for the Bayes factor.

For \emph{disjoint} models (in our case, a model consisting of a Hellings--Downs-correlated red process plus pulsar noise, vs.\ a model consisting of a common-amplitude, spatially-uncorrelated red process plus pulsar noise) we use a \emph{product-space} method \citep{cc95,go01,hhhl16}. In this method we define a super-model that contains all parameters from all models under consideration, as well as an additional model-indexing variable that determines which model is ``active'' and used to evaluate the likelihood.\footnote{This variable is technically discrete, but it can be sampled continuously and cast to an integer to choose the active model.} (In our example, where the parameters are actually the same in both models, the index variable would simply toggle Hellings--Downs correlations in the evaluation of the likelihood.) The ratios of posterior probabilities for two model indices approximate the corresponding Bayes factor. We follow \citet{cl15} to estimate Bayes-factor uncertainties. 

Evaluating the multi-pulsar likelihood is very computationally expensive when we account for inter-pulsar spatial correlations. In that case, we accelerate inference by running at least ten parallel copies of each spatially correlated analysis. These subchains can then be concatenated to form a much larger chain. Each subchain is analyzed to determine that it has ``burned in''\footnote{In MCMC analysis, some early sampled points must be disregarded before the chain can be considered to be sampling from the true posterior probability distribution. The disregarded early portion of the chain is called the ``burn in'' stage.} before combining it with others. To derive upper limits and Savage--Dickey Bayes factors, we simply append the subchains together and proceed as described above. For product-space Bayes factors, we obtain the factor itself from the combined subchains, but we estimate uncertainties in each subchain separately, then add them in quadrature \citep{cl15}.

Arbitrary rules of thumb have been given to interpret the statistical significance of Bayes factors of different magnitudes (see, e.g., \citealt{j61,kr95}), but it is hard to find agreement beyond the trivial statement that factors $\sim 1$ are inconclusive, while very large or small factors point to a strong preference for either model.
In the context of a detection scheme, it seems appropriate to examine the frequentist distribution of the Bayes factor, and to set detection thresholds as a function of false-alarm probability \citep{m12}.
The \emph{sky-scramble} and \emph{phase shifts} methods \citep{tlb+17a,cs16} have been proposed to produce a \emph{background} distribution of the Bayes factor for which spatial correlations are effectively removed from the data.
By contrast, we currently lack a practical approach to establish the significance of a common uncorrelated process; such an approach would likely involve a combination of inference runs on simulated data and cross-validation experiments, such as comparing results for subsets of the dataset.
As we shall see, all the ephemeris-marginalized Bayes factors obtained in this paper are close to unity, and can be deemed inconclusive without a frequentist analysis.

\subsection{Gravitational-wave strain spectrum}
\label{sec:spectralmodels}
The observed timing residuals due to a GWB with characteristic strain $h_c(f)$ are described by the cross-power spectral density
\begin{equation}
    S_{ab}(f) = \Gamma_{ab}(f)\,\frac{{h_c}^2(f)}{12\pi^2\,f^3},
\end{equation}
where $\Gamma_{ab}$ is the overlap reduction function (ORF), which describes correlations between pulsars $a$ and $b$ in the array.
In the case of an isotropic background from SMBHBs the ORF is given by \citet{hd83} (hereafter referred to as H.--D. correlations).
Other correlated effects such as systematic errors in the Solar System ephemeris or clocks can also be described by a timing-residual spectrum that includes a different ORF.

In this paper we consider four models of the GWB spectrum:
\begin{description}
\item[Power-law spectrum]
A population of inspiraling SMBHBs in circular orbits, evolving by GW emission alone produces a characteristic GW-strain spectrum, expressed as
\begin{equation}
h_c(f) = A_\mathrm{GWB} \left( \frac{f}{\mathrm{yr}^{-1}} \right)^{\!\!\alpha}
\label{eq:specdef}
\end{equation}
with $\alpha = -2/3$ \citep{p01}. Different spectral slopes can be used to model relic radiation from the early Universe, under different assumptions for the equation of state of the Universe post-inflation/pre--Big-Bang-Nucleosynthesis (see Sec.\ \ref{sec:primordial}).
We find it expedient to perform our analysis in terms of the timing-residual spectral index $\gamma = 3 - 2\alpha$, such that
\begin{equation}
    S_{ab}(f) = \Gamma_{ab}\,\frac{{A_\mathrm{GWB}}^2}{12\pi^2} \left(\frac{f}{\mathrm{yr}^{-1}}\right)^{\!\!-\gamma}\, \mathrm{yr}^3.
\label{eq:toaspec}
\end{equation}
The fiducial SMBHB $\alpha=-2/3$ then corresponds to $\gamma=13/3$.
\item[Broken--power-law spectrum]
If SMBHBs remain coupled to the dynamics of their galactic environments as they evolve into the nanohertz band, the nanohertz GW strain spectrum will be more complex than described by Eq.\ \eqref{eq:specdef}. This may be the case if three-body scattering of stars from the galactic-center loss cone \citep[e.g.][]{q96,shm06} or interaction with a viscous circumbinary disk \citep[e.g.][]{ks11,hkm09} are a stronger dynamical influence than GW emission at wide orbital separations.
When the binary reaches milliparsec separations, GW emission will always be dominant. 
\citet{scm15} introduced a broken power-law model,
\begin{equation}
h_c(f) = A_\mathrm{GWB} \frac{( f / \mathrm{yr}^{-1} )^\alpha}{\left(1+(f_\mathrm{bend}/f)^\kappa\right)^{1/2}},
\label{eq:broken}
\end{equation}
to model such spectra, where the slope transitions from positive at low frequencies to the canonical $-2/3$ at higher frequencies. The frequency at which the transition occurs encodes information about the typical binary's orbital evolution and astrophysical environment.
\item[Free spectrum]
To characterize the GW-strain sensitivity of our dataset as a function of frequency, we adopt independent uniform priors for the dimensionless-strain amplitudes of each sine--cosine pair of red-process Fourier components (see Sec.\ \ref{sec:ptalike}), corresponding to frequencies $k/T$, with $k = 1, \ldots, N$, where $T$ is the longest timespan in the combined dataset, and $N$ (set to 50 in this paper) is the number of Fourier component pairs. We then derive a joint posterior for all amplitudes.
\item[Gaussian-process spectrum emulation]
This model was introduced by \citet{tss17} as a way to perform searches that are directly informed by detailed source-astrophysics simulations, and to sample the posteriors of the binary environment and dynamics parameters that affect the GW spectrum without generating a new simulation for each likelihood evaluation.
In practice, we perform simulations over a grid in the parameter space of interest, and for each simulation we compute the GW characteristic strain spectrum. We then train a Gaussian process \citep{rw06} to interpolate over all spectra in parameter space, allowing spectral amplitudes to be predicted at any other point with an associated normal uncertainty.
We then use these predictions and uncertainties as priors on the strain amplitude at each frequency within the free-spectrum model.
\end{description}

\subsection{Solar System ephemeris errors and uncertainties}
\label{sec:eph}

A Solar System ephemeris is used in pulsar timing to convert observatory TOAs to an inertial frame centered at the Solar System barycenter, factoring out all effects due to Earth's motion. The dominant correction to the TOAs is the \emph{Roemer delay}---the classical light-travel time between the geocenter and the Solar System barycenter. Pulsar-timing studies have typically relied on the latest SSE released by JPL, adopting it as a \emph{model with fixed parameters}---that is, without including any SSE parameter uncertainties or corrections in timing-model fits.
In the early stages of our analysis of the NANOGrav 11-year dataset, we became aware that the choice of SSE among the latest few released by JPL has a measurable impact on our GWB upper limits and model-comparison Bayes factors. Indeed, the abundance and precision of NANOGrav's measurements are now such that the accuracy to which we can estimate the Earth's orbit around the SSB limits our sensitivity to GWs. SSE errors have been speculated on as a source of potential bias in PTA GW detection efforts \citep{thk+15}, but this paper marks the first time that this effect has been rigorously studied with real datasets.

The JPL SSEs\footnote{\href{https://ssd.jpl.nasa.gov/?ephemerides}{https://ssd.jpl.nasa.gov/?ephemerides}}, as well as the French INPOP\footnote{\href{https://www.imcce.fr/inpop}{https://www.imcce.fr/inpop}}, fit the orbits and masses of a large set of Solar System bodies to a heterogenous dataset collected over the last few decades, using spacecraft ranging, direct planetary radar ranging, spacecraft VLBI, and (for the Moon) laser-ranging of retroreflectors left by the Apollo missions. The orbits are integrated numerically from initial conditions (``epoch'' positions and velocities), which are the parameters that are fit for, together with other quantities such as the masses of minor Solar System bodies [but not planet masses, which are estimated separately from observed motions in planetary systems \citep{2009IPNPR.178C...1F}]. The resulting SSEs are distributed as Chebyshev polynomials over a range of dates; notably, they do not include estimates of orbit uncertainties and of possible systematics.

To investigate the effects of SSE errors, we repeated all upper-limit and model-comparison analyses in this paper using the four most recent JPL SSEs [DE421, released in 2008 \citep{2009IPNPR.178C...1F}; DE430 \citep{2014IPNPR.196C...1F}; DE435 \citep{de435}; DE436 \citep{de436}]; for the simplest analysis, we used also the French INPOP$13$c \citep{INPOP13c}.
The orbit of Earth relative to the Sun is consistent at the $10$-m level across these ephemerides, after accounting for an overall rotation w.r.t.\ the International Celestial Reference Frame, which originates from updated very-long-baseline-interferometry observations of spacecraft at Mars. However, the orbit of the Sun w.r.t.\ the SSB and (therefore) the orbit of Earth w.r.t.\ the SSB match only at the $100$-m level. This discrepancy is attributed to differences in the estimated masses and positions of Jupiter, Uranus, and Neptune.
Hence, our GW analysis shows significant systematic differences among the upper limits and Bayes factors computed using different ephemerides. Near-future efforts may lead to improvements in the ephemeris accuracy that are appropriate for pulsar timing, namely: $(i)$ estimates of Jupiter's orbit will be improved by including \textit{Juno} spacecraft data in the SSE fit; $(ii)$ ranging data from \textit{Cassini} may better estimate the mass of Uranus; $(iii)$ \textit{Gaia} data may improve orbit estimates for Uranus and Neptune; and finally $(iv)$ pulsar-timing data may be used to improve the estimate of Neptune's mass.

We account for the differences between SSEs by developing a physical model (\bayesephem) that corrects Earth's tabulated orbit using eleven parameters.
Four of these correspond to perturbations in the masses of the outer planets, and generate corrections $-(\delta M_i/M_\mathrm{tot}) \, \mathbf{r}_i(t)$, where $\delta M_i$ is the outer-planet's mass correction, $M_\mathrm{tot}$ is the total mass of the Solar System, and $\mathbf{r}_i(t)$ is the outer-planet's orbit.
One parameter describes a rotation rate about the ecliptic pole: this accounts for differences in the estimated semi-major axis of the Earth--Moon-barycenter orbit, which gives rise to a linear rate in estimated ecliptic longitude.
Since the orbit of the Sun about the SSB is largely influenced by Jupiter, and since the Jovian period is comparable to the span of our dataset, we also include corrections to Earth's orbit generated by perturbing Jupiter's average orbital elements. These corrections have the form $-(M_J/M_\mathrm{tot}) \, (\partial \mathbf{r}_J(t)/\partial a_J^\mu) \, \delta a_J^\mu$, where the partial derivatives encode the changes in Jupiter's orbit as we change the orbital elements, and where the six $\delta a_J^\mu$ are the orbital-element perturbations (which we define using Brouwer and Clemence's (\citeyear{1961mcm..book.....B}) ``set-III'' parameters).
By contrast, Saturn's orbit is constrained more strongly by available data, while Uranus and Neptune have large orbit uncertainties but very long periods, so they can only generate linear-in-time Roemer biases that are absorbed by fitting pulsar periods.

Thus, we present GW upper limits and model-comparison Bayes factors that are marginalized over these SSE uncertainty parameters. We regard these \bayesephem\ limits and Bayes factors as our fiducial results in this paper.
To derive them, we constrain the outer-planet masses using the current IAU best estimates \citep{iaumasses,jh+2000,2006AJ....132.2520J,2014AJ....148...76J,2009AJ....137.4322J}, and use IAU uncertainties to set Gaussian priors.
The rate of rotation about the ecliptic pole is left unconstrained.
We experimented with setting priors for Jupiter's orbital elements using estimated uncertainties,\footnote{\citet{FPP17} estimate uncertainties in Jupiter and Saturn orbits by comparing fits that use independent subsets of the data for each planet.} but we find better results using uninformative priors.
This is not surprising, because Jupiter's orbital elements are highly correlated with those of the other planets, and our linearized correction of Jupiter's orbit cannot account for those correlations.
Nevertheless, the resulting variations of Earth's orbit are comparable with the systematic differences that we observe across JPL SSEs, which we take as evidence that the \bayesephem\ uncertainty parameters are representative of true SSE uncertainties.

Our Bayesian-inference studies produce orbital-element posteriors for Jupiter corresponding to position offsets at the level of $\sim 100\;\mathrm{km}$. We defer the full details of our investigations of SSE uncertainties and systematics to an upcoming paper, where we compare reconstructed Jupiter orbits from our analysis to those from the JPL ephemerides, and discuss potential modeling improvements to \textsc{BayesEphem}.

\subsection{Data model and likelihood}
\label{sec:ptalike}

Except for Gaussian-process spectrum emulation and for the treatment of SSE errors, the data model used in this paper matches that of \citetalias{abb+16} closely, so we refer the reader to that publication for an overview of noise modeling, marginalization over timing-model parameters, our rank-reduced formalism for time-correlated processes (e.g., timing noise or GWB), and the PTA likelihood.

The rank-reduced formalism refers to the expansion of processes on a sine--cosine Fourier basis with frequencies $k/T$, where $T$ is the span between the minimum and maximum TOA in the array. The number of basis vectors is chosen to be high enough that inference results are insensitive to adding more: we use 30 for all applications except for the free-spectrum GWB model, for which we use 50.

As for the PTA likelihood, we introduced a significant change compared to \citetalias{abb+16}. ``ECORR'' (jitter-like) noise is fully correlated for simultaneous observations at different observing frequencies, but fully uncorrelated in time. In \citetalias{abb+16}, we treated ECORR degrees of freedom by assigning them ``exploder'' basis vectors, and then analytically marginalizing their coefficients simultaneously with timing-model, red-noise, and GWB coefficients. Doing so becomes computationally prohibitive when H.--D.\ correlations are included. In this paper, we include ECORR noise as block-diagonal entries (one block per epoch per backend--receiver system) in the otherwise diagonal white-noise covariance matrix, and invert the matrix using the fast \citet{sm49} formula. Doing so eliminates a significant computational bottleneck.

As in \citetalias{abb+16}, computational efficiency is also helped by fixing all white-noise parameters to their 1D maximum \emph{a posteriori} values from single-pulsar noise studies. This choice is justified empirically by the very small variance of white-noise parameters. 


\newcommand{\cc}{\checkmark}
\begin{table}[t]
\begin{center}
\label{tab:spatialcorr_modeltab}
\caption{Spatially correlated red-noise processes used in our analysis. All models include intrinsic white-noise and red-noise processes in each pulsar; additional common processes (with the same characteristic amplitude and spectrum in every pulsar) can be uncorrelated, or have Hellings--Downs (GW-like), dipolar (ephemeris-error--like), and monopolar (clock-error--like) spatial correlations. Model 2A (uncorrelated common process) was used to derive the main results of \citetalias{abb+16}; model 3A (Helling--Downs-correlated common process) is the fiducial model used to constrain the GWB in this publication.}
\begin{tabular}{lc@{\;\;\;}c@{\;\;\;}c@{\;\;\;}c@{\;\;\;}c@{\;\;\;}c@{\;\;\;}c@{\;\;\;}c@{\;\;\;}c}
\hline \hline
 & \multicolumn{9}{c}{model} \\
red-noise process & 1 & 2A & 2B & 2C & 2D & 3A & 3B & 3C & 3D \\
\hline
intrinsic (per pulsar) &\cc& \cc& \cc& \cc& \cc& \cc& \cc& \cc& \cc\\
uncorr. common 		&   & \cc&    &    &    &    &    &    &    \\
H.-D. corr. common 				&   &    &    &    &    & \cc& \cc& \cc& \cc\\
dipole corr. common 				&   &    & \cc& \cc&    &    & \cc& \cc&    \\
monopole corr. common 			&   &    &    & \cc& \cc&    &    & \cc& \cc\\
\hline \hline
\end{tabular}
\end{center}
\end{table}


\begin{table*}[ht]
\begin{center}
\scriptsize
\caption{Prior distributions used in all analyses performed in this paper.}
\label{tab:priors}
\begin{tabular}{llll}
\hline\hline
parameter & description & prior & comments \\
\hline
\multicolumn{4}{c}{White Noise} \\[1pt]
$E_{k}$ & EFAC per backend/receiver system & Uniform $[0, 10]$ & single-pulsar analysis only \\
$Q_{k}$ [s] & EQUAD per backend/receiver system & log-Uniform $[-8.5, -5]$ & single-pulsar analysis only \\
$J_{k}$ [s] & ECORR per backend/receiver system & log-Uniform $[-8.5, -5]$ & single-pulsar analysis only \\
\hline

\multicolumn{4}{c}{Red Noise} \\[1pt]
$A_{\rm red}$ & red-noise power-law amplitude & Uniform $[10^{-20}, 10^{-11}]$ (upper limits) & \\
& & log-Uniform $[-20, -11]$ (model comparison) & one parameter per pulsar  \\
$\gamma_{\rm red}$ & red-noise power-law spectral index & Uniform $[0, 7]$ & one parameter per pulsar \\
\hline

\multicolumn{4}{c}{\textsc{BayesEphem}} \\[1pt]
$z_{\rm drift}$ [rad/yr] & drift-rate of Earth's orbit about ecliptic $z$-axis & Uniform [$-10^{-9}, 10^{-9}$] & one parameter for PTA \\
$\Delta M_{\rm jupiter}$ [$M_{\odot}$] & perturbation to Jupiter's mass & $\mathcal{N}(0, 1.55\times 10^{-11})$  & one parameter for PTA \\
$\Delta M_{\rm saturn}$ [$M_{\odot}$] & perturbation to Saturn's mass & $\mathcal{N}(0, 8.17\times 10^{-12})$  & one parameter for PTA \\
$\Delta M_{\rm uranus}$ [$M_{\odot}$] & perturbation to Uranus' mass & $\mathcal{N}(0, 5.72\times 10^{-11})$  & one parameter for PTA \\
$\Delta M_{\rm neptune}$ [$M_{\odot}$] & perturbation to Neptune's mass & $\mathcal{N}(0, 7.96\times 10^{-11})$  & one parameter for PTA \\
PCA$_{i}$ & $i$th PCA component of Jupiter's orbit & Uniform $[-0.05, 0.05]$ & six parameters for PTA \\
\hline

\multicolumn{4}{c}{Monopole-correlated clock-error signal, power-law spectrum} \\[1pt]
$A_{\rm mono}$ & Equivalent strain amplitude & Uniform $[10^{-18}, 10^{-11}]$ (upper limits) & \\
& & log-Uniform $[-18, -14]$ (model comp., $\gamma=13/3$) & one parameter for PTA \\
& & log-Uniform $[-18, -11]$ (model comp., $\gamma$ varied) & one parameter for PTA \\
$\gamma_\mathrm{mono}$ & GWB power-law spectral index & delta function  & fixed, depends on analysis \\
\hline

\multicolumn{4}{c}{Dipole-correlated SSE-error signal, power-law spectrum} \\[1pt]
$A_{\rm dip}$ & Equivalent strain amplitude & Uniform $[10^{-18}, 10^{-11}]$ (upper limits) & \\
& & log-Uniform $[-18, -14]$ (model comp., $\gamma=13/3$) & one parameter for PTA \\
& & log-Uniform $[-18, -11]$ (model comp., $\gamma$ varied) & one parameter for PTA \\
$\gamma_\mathrm{dip}$ & GWB power-law spectral index & delta function  & fixed, depends on analysis \\
\hline

\multicolumn{4}{c}{GWB, power-law spectrum} \\[1pt]
$A_{\rm GWB}$ & GWB strain amplitude & Uniform $[10^{-18}, 10^{-11}]$ (upper limits) & \\
& & log-Uniform $[-18, -14]$ (model comp., $\gamma_\mathrm{GWB}=13/3$) & one parameter for PTA \\
& & log-Uniform $[-18, -11]$ (model comp., $\gamma_\mathrm{GWB}$ varied) & one parameter for PTA \\
$\gamma_{\rm GWB}$ & GWB power-law spectral index & delta function  & fixed, depends on analysis \\
\hline

\multicolumn{4}{c}{GWB, free spectrum} \\[1pt]
$\rho_{i}$ [s$^{2}$] & GWB power-spectrum coefficients at $f=i/T$ & uniform in $\rho_{i}^{1/2}$ $[10^{-18},10^{-8}]^{a}$ & one parameter per frequency\\
\hline

\multicolumn{4}{c}{GWB, broken--power-law spectrum} \\[1pt]
$A_{\rm GWB}$ & GWB broken power-law amplitude & log-Normal   & one parameter for PTA \\
& & $\mathcal{N}(-14.4, 0.26)$ & \citetalias{mop14} \\
& & $\mathcal{N}(-15, 0.22)$  & \citetalias{s13} \\
& & $\mathcal{N}(-14.95, 0.12)$ & \citet{ss16}$^{b}$  \\
& & $\mathcal{N}(-14.82, 0.08)$ & \citet{ss16}$^{c}$  \\
$\gamma_{\rm GWB}$ & GWB power-law spectral index & delta function  & fixed to $13/3$\\
$\kappa$ & GWB broken power-law low-freq.\ spectral index & Uniform [0,7] & one parameter for PTA \\
$f_{\rm bend}$ [Hz] & GWB broken power-law bend frequency & log-Uniform  [$-9$,$-7$] & one parameter for PTA \\
\hline

\multicolumn{4}{c}{GWB, Gaussian-process--interpolated spectrum} \\
$\rho_{i}$ [s$^{2}$] & GWB power-spectrum coefficients at $f = i/T$ & $\mathcal{N}(0, V(\alpha_{\rm BH}, \rho_{\rm stars}, e_{0}))$ & one parameter per frequency \\
$\alpha_{\rm BH}$ & $y$-intercept of $M_\mathrm{BH}-M_{\rm bulge}$ relation & Uniform $[7, 9]$ & one parameter for PTA \\
$\rho_{\rm stars}$ [$M_{\odot}\mathrm{pc}^{-3}$] & mass density of galactic-core stars & log-Uniform $[1, 4]$ & one parameter for PTA \\
$e_{0}$ & binary eccentricity at formation & Uniform [0, 0.95] & one parameter for PTA \\
\hline
\end{tabular}
{\flushleft
{$^{a}$ The uniform $\rho_{i}^{1/2}$ prior is chosen to be consistent with the uniform $A_\mathrm{GWB}$ prior for the power-law model, since $\varphi_{ii}\propto A_\mathrm{GWB}^{2}$.} \\
{$^{b}$ Uses  \cite{mm13} $M_\mathrm{BH}-M_{\rm bulge}$ relation.} \\
{$^{c}$ Uses  \cite{kh13} $M_\mathrm{BH}-M_{\rm bulge}$ relation.} \\
}
\end{center}
\end{table*}

Our upper-limit and model-comparison studies are performed under a variety of assumptions about the presence of red-spectrum processes: in addition to individual red-spectrum timing noise for each pulsar, we model the GWB as a spatially uncorrelated common process (a computational simplification appropriate in the weak-GWB limit, used in \citetalias{abb+16}) and as a Hellings--Downs-correlated common process (our fiducial GWB model); we also consider common processes with different correlations (dipolar, as appropriate for SSE errors, and monopolar, as appropriate for long-timescale clock errors). \autoref{tab:spatialcorr_modeltab} describes the nine models used in this paper, which are labeled 1, 2A--D, and 3A--D. In model-class $1$ only intrinsic pulsar noise processes are included; in model-class $2$ there are intrinsic pulsar noise processes, as well as non-GW noise processes that induce inter-pulsar spatial correlations (such as clock and SSE errors); in model-class $3$ we include a GWB signal. The roman characters given after the model-class number indicate the specific combination of noise and signal processes forming the model. 

We perform each analysis by adopting each of the DE421, DE430, DE435, and DE436 (and occasionally INPOP$13$c) ephemerides as fixed-parameter models, and by marginalizing over SSE uncertainties using \bayesephem. Our Bayesian priors for all parameters are described in \autoref{tab:priors}.

\subsection{Optimal Statistic}
\label{sec:optimalstatdef}

As in \citetalias{abb+16}, we perform a frequentist GWB analysis using the \emph{optimal statistic} $\hat{A}_\mathrm{GWB}^2$, a point estimator for the amplitude of an isotropic GW stochastic background \citep{abc+09,ccs+15}. This statistic accounts implicitly for inter-pulsar spatial correlations. The estimator is derived by maximizing the PTA likelihood analytically, and it can be written as
\begin{equation}
	\hat{A}_\mathrm{GWB}^2 = \frac{\sum_{ab} {\delta\bf{t}}_a^T {\bf{P}}_a^{-1} \tilde{{\bf{S}}}_{ab} {\bf{P}}_b^{-1} {\delta\bf{t}}_b}{\sum_{ab} \Tr \left( {\bf{P}}_a^{-1} \tilde{{\bf{S}}}_{ab} {\bf{P}}_b^{-1} \tilde{{\bf{S}}}_{ba} \right) } \,,
\end{equation}
where ${\delta \bf{t}}_a$ is the vector of timing residuals for pulsar $a$, ${\bf{P}}_a = \left\langle {\delta\bf{t}}_a {\delta\bf{t}}_a^T \right\rangle$ is the autocovariance matrix of the residuals, and $\hat{A}_\mathrm{gw}^2 \tilde{{\bf{S}}}_{ab} = {\bf{S}}_{ab} = \left. \left\langle {\delta\bf{t}}_a {\delta\bf{t}}_b^T \right\rangle \right|_{a \neq b}$ is the cross-covariance matrix between the residuals for pulsars $a$ and $b$. 
The average signal-to-noise ratio (SNR) of the optimal statistic is
\begin{equation}
	\left\langle \rho \right\rangle = A_\mathrm{gw}^2 \left[ \sum_{ab} \Tr \left( {\bf{P}}_a^{-1} \tilde{{\bf{S}}}_{ab} {\bf{P}}_b^{-1} \tilde{{\bf{S}}}_{ba} \right) \right]^{1/2} \,,
\end{equation}
which is a measure of the significance of inter-pulsar spatial correlations. When drawing comparisons between results produced using this frequentist technique and our Bayesian techniques, the relevant model selection is between models $3$A and $2$A.

We use two procedures to compute $\hat{A}_\mathrm{GWB}^2$. 
In the more conventional \emph{fixed-noise} analysis, 
we compute $\hat{A}_\mathrm{GWB}^2$ 
at fixed values of the pulsar red-noise parameters $A_\mathrm{red}$ and $\gamma_\mathrm{red}$. 
The red-noise parameters are the values that jointly maximize the likelihood, 
as found in a Bayesian parameter-estimation study 
that includes pulsar red-noise and a common red-noise process. 
In the newer \emph{noise-marginalized} analysis \citep{viet17}, 
we use posterior samples from a Bayesian study 
to marginalize the optimal statistic over pulsar red-noise parameters. 
This results in distributions for both $\hat{A}_\mathrm{GWB}^2$ and the SNR, 
rather than a single value of $\hat{A}_\mathrm{GWB}^2$ and a corresponding SNR. 
In both cases, pulsar white-noise parameters are fixed to their maximum-likelihood values, as determined individually for each pulsar with Bayesian inference. 
As discussed in \citet{viet17}, simulations show that the noise-marginalized technique produces more accurate estimates of $A_\mathrm{GWB}$ compared to the fixed-noise technique. This is because the pulsar red-noise parameters are highly covariant with common-process red-noise parameters, so the fixed-noise analysis tends to systematically underestimate the amplitude and significance of common signals.

\subsection{Software}
\label{sec:software}

We generated most of the results in this paper using the open-source software package \texttt{NX01}\footnote{\href{https://github.com/stevertaylor/NX01}{https://github.com/stevertaylor/NX01}} \citep{nx01}, which implements the PTA likelihood and priors. \texttt{NX01} was validated on a wide range of problems, including several 11-year analyses, by cross-comparison with the well-established  \texttt{PAL2}\footnote{\href{https://github.com/jellis18/PAL2}{https://github.com/jellis18/PAL2}} \citep{evh17a} and with NANOGrav's new flagship package, \texttt{enterprise}\footnote{\href{https://github.com/nanograv/enterprise}{https://github.com/nanograv/enterprise}} \citep{enterprise}.
We perform MCMC using \texttt{PTMCMCSampler}\footnote{\href{https://github.com/jellis18/PTMCMCSampler}{https://github.com/jellis18/PTMCMCSampler}} \citep{evh17b}, which implements a variety of proposal schemes (adaptive Metropolis, differential evolution, parallel tempering, etc.), which can be used together in the same run.

As a companion to this paper, we are releasing a \texttt{Docker}\footnote{\href{https://github.com/nanograv/11yr_stochastic_analysis}{https://github.com/nanograv/11yr\_stochastic\_analysis}} image that contains a full stack of our software (including all required libraries), and that can be used to reproduce the upper limits, Bayes factors, as well as many of the figures of this paper, using \texttt{enterprise}.

\section{Results}
\label{sec:results}

All results in this paper are based on a subset of the full 11-year data release, which includes the 34 pulsars with a timing baseline greater than 3 years. This restriction is justifiable since we do not expect any detectable GW signal to be present at frequencies $\gtrsim 3 \mathrm{yr}^{-1}$, and it has the advantage of making our spatially correlated analysis -- required to search for Hellings--Down correlations in the residuals -- more computationally tractable, since the computational cost scales roughly as the cube of the number of pulsars.
\autoref{tab:pulsar} lists the 34 pulsars with their epoch-averaged RMS residuals, number of epochs and TOAs, and timing baselines.
\begin{table}[t]
\begin{center}
\caption{\label{tab:pulsar} Pulsars used in our GWB analysis (see \citetalias{abb+17} for full details of pulsar properties). The second column shows the weighted root-mean-square epoch-averaged post-fit timing residuals (see \citetalias{abb+15} for a definition of this quantity).}

\begin{tabular}{ldddd}
\hline\hline
PSR name &
\multicolumn{1}{r}{RMS [$\mu$s]} &
\multicolumn{1}{l}{\#epochs} &
\multicolumn{1}{l}{\#TOAs} &
\multicolumn{1}{l}{baseline [yr]} \\
\hline
J0023$+$0923 & 0.361 & 415 & 8,217 & 4.4  \\
J0030$+$0451 & 0.691 & 268 & 5,699 & 10.9  \\
J0340$+$4130 & 0.454 & 127 & 6,475 & 3.8  \\
J0613$-$0200 & 0.422 & 324 & 11,566 & 10.8  \\
J0645$+$5158 & 0.178 & 166 & 6,370 & 4.5  \\
J1012$+$5307 & 1.07 & 493 & 16,782 & 11.4  \\
J1024$-$0719 & 0.323 & 194 & 8,233 & 6.2  \\
J1455$-$3330 & 0.672 & 277 & 7,526 & 11.4  \\
J1600$-$3053 & 0.23 & 275 & 12,433 & 8.1  \\
J1614$-$2230 & 0.199 & 241 & 11,173 & 7.2  \\
J1640$+$2224 & 0.426 & 323 & 5,982 & 11.1  \\
J1643$-$1224 & 3.31 & 298 & 11,528 & 11.2  \\
J1713$+$0747 & 0.108 & 789 & 27,571 & 10.9  \\
J1738$+$0333 & 0.52 & 208 & 4,881 & 6.1  \\
J1741$+$1351 & 0.128 & 134 & 3,047 & 6.4  \\
J1744$-$1134 & 0.842 & 322 & 11,550 & 11.4  \\
J1747$-$4036 & 3.59 & 113 & 6,065 & 3.8  \\
J1853$+$1303 & 0.239 & 107 & 2,514 & 4.5  \\
B1855$+$09 & 0.809 & 296 & 5,634 & 11.0  \\
J1903$+$0327 & 3.65 & 112 & 3,326 & 6.1  \\
J1909$-$3744 & 0.148 & 451 & 17,373 & 11.2  \\
J1910$+$1256 & 0.544 & 130 & 3,563 & 6.8  \\
J1918$-$0642 & 0.322 & 364 & 12,505 & 11.2  \\
J1923$+$2515 & 0.229 & 87 & 1,954 & 4.3  \\
B1937$+$21 & 1.57 & 460 & 14,217 & 11.3  \\
J1944$+$0907 & 0.352 & 104 & 2,850 & 4.4  \\
B1953$+$29 & 0.377 & 88 & 2,331 & 4.4  \\
J2010$-$1323 & 0.257 & 222 & 10,844 & 6.2  \\
J2017$+$0603 & 0.11 & 102 & 2,359 & 3.8  \\
J2043$+$1711 & 0.12 & 197 & 3,262 & 4.5  \\
J2145$-$0750 & 0.968 & 258 & 10,938 & 11.3  \\
J2214$+$3000 & 1.33 & 176 & 4,569 & 4.2  \\
J2302$+$4442 & 1.07 & 138 & 6,549 & 3.8  \\
J2317$+$1439 & 0.271 & 395 & 5,958 & 11.0  \\
\hline \hline
\end{tabular}

\end{center}
\end{table}

As discussed in Sec.\ \ref{sec:ptalike}, we perform analyses for variants of our data model that reflect different assumptions about common red-spectrum processes, as listed in \autoref{tab:spatialcorr_modeltab}, and under four JPL ephemerides as well as \bayesephem\ (in select cases we include also the French INPOP13, which yields results broadly similar to DE430).

\subsection{Bayesian upper limits}
\label{sec:balimit}

Following \citetalias{abb+16}, we present upper limits on the strain amplitude of a GWB modeled as a power law and as a free spectrum (see Sec.\ \ref{sec:spectralmodels}).


\begin{table*}[ht]
	\begin{center}
	\caption{GWB-amplitude 95\% upper limits for the NANOGrav 11-year dataset, computed for a power-law spectrum with $\gamma = 13/3$, and with uniform prior on $A_\mathrm{GWB}$ [see Eqs.\ \eqref{eq:specdef} and \eqref{eq:toaspec}]. We report limits for an uncorrelated common process (as in \citetalias{abb+16}), and for a Hellings--Downs spatially correlated process, either alone (in bold, our fiducial result) or in the presence of additional correlated processes with different ORF.}
	\begin{tabular}{@{} lccccc @{}}
		\hline\hline
    		\multirow{3}{*}{JPL ephemeris}	& \multicolumn{4}{c}{$95\%$ upper limit on $A_\mathrm{GWB}$ [$\times 10^{-15}$], $\gamma = 13/3$ power law} \\
								& uncorrelated common process (2A) & \multicolumn{4}{c}{H.--D.\ correlated common process} \\					
								&  & alone (3A) & + dipole (3B) & + monopole, dipole (3C) & + monopole (3D) \\					
    		\hline
    		DE421 				& $1.505(8)$ & $\mathbf{1.53(1)}$ & $1.478(8)$ & $1.487(8)$ & $1.53(3)$\\
    		DE430 				& $1.76(2)\phantom{0}$ 	 & $\mathbf{1.79(1)}$ & $1.698(9)$ & $1.676(9)$ & $1.74(2)$\\
    		DE435 				& $1.57(3)\phantom{0}$ 	 & $\mathbf{1.60(1)}$ & $1.555(8)$ & $1.55(1)\phantom{0}$ & $1.58(2)$\\
    		DE436 				& $1.61(2)\phantom{0}$ 	 & $\mathbf{1.67(1)}$ & $1.594(9)$ & $1.56(1)\phantom{0}$ & $1.60(2)$\\
		INPOP13c 			    & $1.74(3)\phantom{0}$ 	& 	--- 		& --- 		& --- & ---\\
		\hline
		\textsc{BayesEphem} 	& $1.34(1)\phantom{0}$ 	& $\mathbf{1.45(2)}$ & $1.52(3)\phantom{0}$ & $1.49(3)\phantom{0}$ & $1.48(4)$\\
    		\hline
	\end{tabular}
	\end{center}
	\label{tab:bayes_uls}
\end{table*}

We first discuss our limit on the GWB from a population of GW-driven SMBHB inspirals, as represented by Eq.\ \eqref{eq:toaspec} for $\gamma = 13/3$.
Adopting each of the JPL ephemerides as fixed-parameter models, and including Hellings--Downs inter-pulsar correlations in the likelihood (model 3A), the $95\%$ upper limit on $A_\mathrm{GWB}$ ranges from $1.53(1) \times 10^{-15}$ for DE421 to $1.78(1) \times 10^{-15}$ for DE430, where the digits in parentheses give 1-$\sigma$ uncertainties (see the third bolded column of \autoref{tab:bayes_uls}). Indeed, the choice of ephemeris leads to systematic biases that are larger than the statistical uncertainty of the limits. Furthermore, the limits do not evolve monotonically with later ephemerides, suggesting that even DE436, the most recent ephemeris based on the most data, is still measurably displaced from truth for the purposes of our analysis.

We therefore chose to marginalize the $A_\mathrm{GWB}$ posterior over ephemeris uncertainties, using the \bayesephem\ model described in Sec.\ \ref{sec:eph}. Doing so yields our \emph{fiducial 11-year upper limit} of $1.45(2) \times 10^{-15}$. This value is the same, within sampling error, no matter which ephemeris we take as a starting point for \bayesephem, demonstrating that we have successfully ``bridged'' the individual ephemerides.

Comparing the columns of \autoref{tab:bayes_uls} shows how the upper limits vary under different assumptions on the presence of spatially correlated common processes in the data. The limits are slightly more stringent if we model the GWB as a spatially uncorrelated common process (model 2A in the second column), indicating that Hellings--Downs correlations help the likelihood isolate a GW-like signal (whether real, or due to random noise fluctuations). Introducing additional spatially correlated processes (with ephemeris-error--like dipolar correlations, clock-error--like monopolar correlations, or both, corresponding to models 3B, 3D, and 3C) reduces upper limits for the individual ephemerides but not for \bayesephem, suggesting that the same realization of inter-pulsar signal correlations can be picked up by different ORFs, and that dipole and monopole processes can absorb some, but not all, of the systematic bias caused by ephemeris error.

\begin{figure}
  \includegraphics[width=\columnwidth]{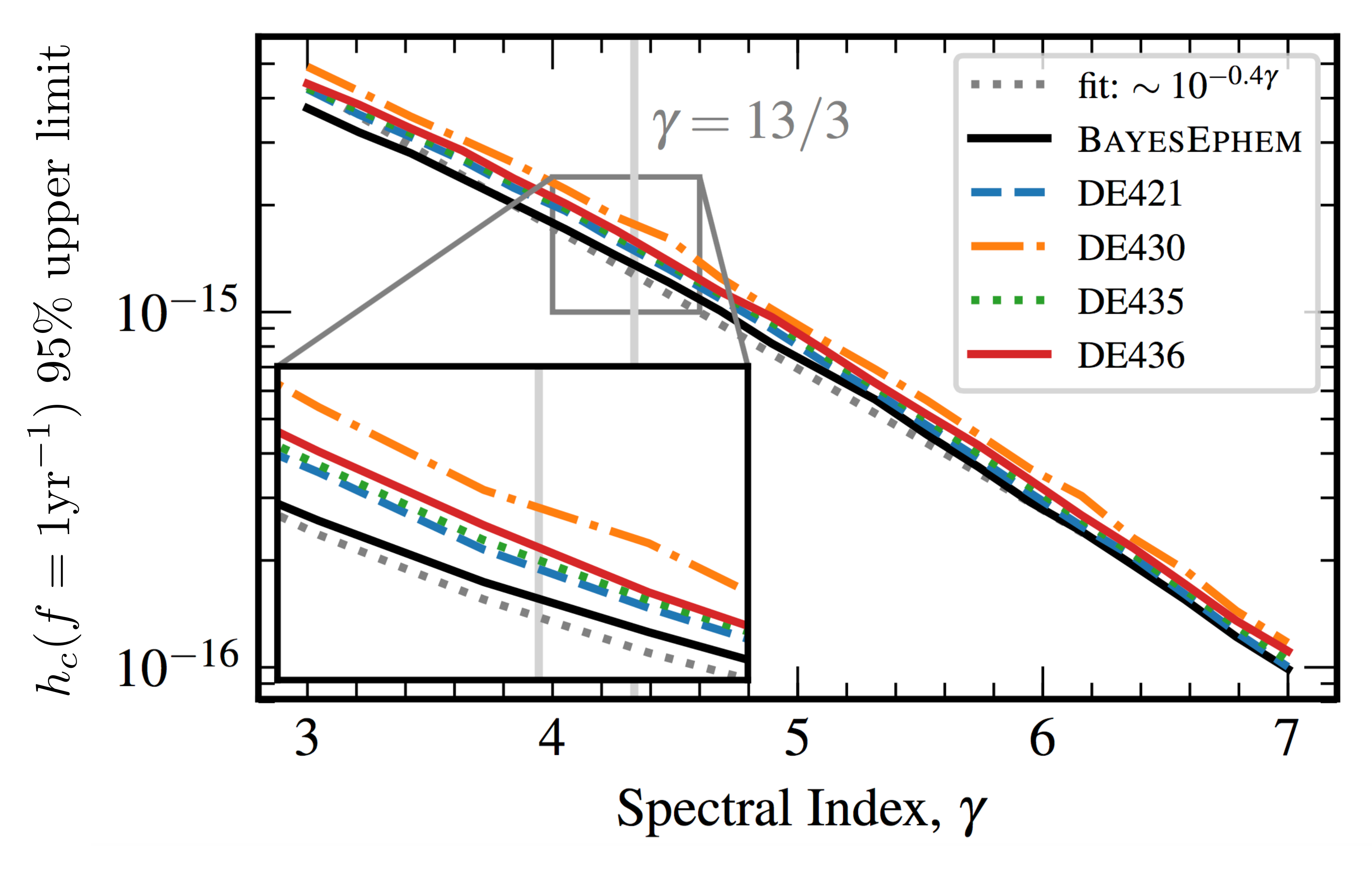}
  \caption{GWB-amplitude 95\% upper limit for an uncorrelated common process (model 2A) as a function of spectral index $\gamma$ (see Eq.\ \eqref{eq:toaspec}), for the JPL ephemerides and for \bayesephem. The dotted curve shows a power-law fit to the \bayesephem\ curve, which is consistent with a similar fit in \citetalias{abb+16}.}
  \label{fig:Agw_v_gamma}
\end{figure}
In \autoref{fig:Agw_v_gamma} we show the 95\% upper limit for the amplitude of an uncorrelated common process  (model 2A) as a function of $\gamma$. In the absence of red noise, and if the lowest sampling frequency ($1/T$) dominated our sensitivity, we would expect these constraints to scale as $\propto T^{-\gamma/2}$, where $T$ is the longest timing baseline across the entire PTA.
We find the actual scaling to be closer to $\propto T^{-0.4\gamma}$, indicating that red noise is present and that more than one frequency component contributes to the likelihood.

\begin{figure}
  \includegraphics[width=\columnwidth]{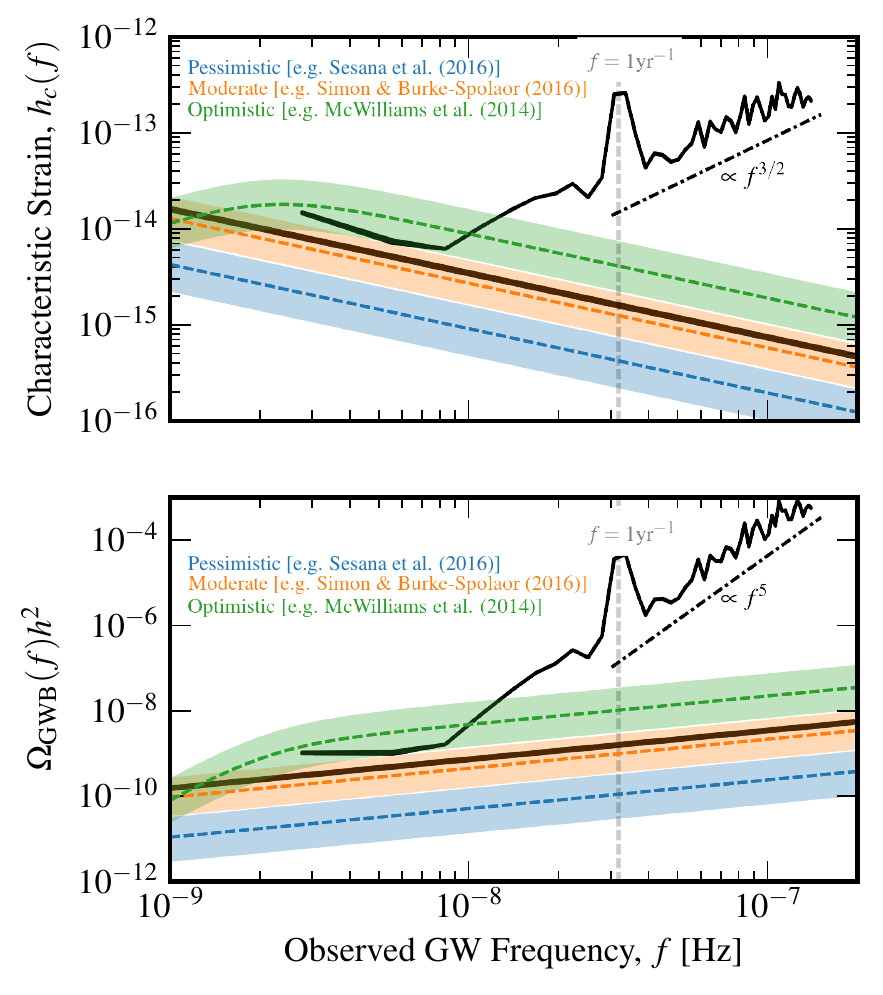} \\
  \caption{\textit{Top panel:} GWB-amplitude 95\% upper limits for an uncorrelated common process with $\gamma = 13/3$ power law (straight black line) or with independently determined free-spectrum components (jagged black line). The thickness of the lines spans the spread of results over different ephemerides. The dash-dotted line shows the expected sensitivity scaling behavior for white-noise.
The colored dashed lines and bands show median and one-sigma ranges for the GWB amplitudes predicted in \citetalias{mop14} (green), \citet{ss16} (orange), and \citetalias{s16} (blue). \textit{Bottom panel:} As in the top panel, except showing the results in terms of the stochastic GWB energy density (per logarithmic frequency bin)in the Universe as a fraction of closure density, $\Omega_\mathrm{GWB}(f)h^2$. The relationship between $h_c(f)$ and $\Omega_\mathrm{GWB}(f)h^2$ is given in \autoref{eq:omega_gwb}.}
  \label{fig:hcf}
\end{figure}
In the top panel of \autoref{fig:hcf} we show 95\% upper limits for free-spectrum amplitudes (jagged black line), which are diagnostic of the sensitivity of our dataset to individual monochromatic GW signals. In the same plot we show also the $\gamma = 13/3$ ($\alpha=-2/3$) power law limit (straight black line). The thickness of the lines indicates the spread of results over ephemeris choices.
Sensitivity is inhibited at lower frequencies by fitting pulsar spindown parameters, and is dominated at higher frequencies by white noise, matching the expected $f^{3/2}$ expected slope shown as the dash-dotted line. Sensitivity is also noticeably lost at $f = \mathrm{yr}^{-1}$ by fitting pulsar positions.
The colored dashed lines and bands display a representative selection of theoretical expectations for the SMBHB GWB: \citet{mop14} (hereafter \citetalias{mop14}); \citet{ss16}, and \citet{s16} (hereafter \citetalias{s16}). The models in \citet{ss16} and \citetalias{s16} are broadly similar, differing predominantly in the choice of SMBH--host-galaxy mass relationship, where \citetalias{s16} factors in potential selection biases in dynamically-measured SMBH masses \citep{sbs+16}. These same results and theoretical expectations are shown in the bottom panel of \autoref{fig:hcf} in terms of the stochastic GWB energy density (per logarithmic frequency bin) in the Universe as a fraction of closure density, $\Omega_\mathrm{GWB}(f)h^2$, where the scaling by $h^2$ makes the results agnostic of the specific value of the Hubble constant. The fractional energy density scales as $\Omega_\mathrm{GWB}h^2 \propto f^2h_c(f)^2$. In Sec.\ \ref{sec:smbhb} below we present an astrophysical discussion of our results.

\subsection{Bayesian model-comparison evidence for GWs}
\label{sec:bamodelcomp}

\begin{table*}[ht]
\begin{center}
    \scriptsize
    \label{tab:table_sd_bfs}
    \caption{Bayes factors for model comparisons using NANOGrav's $11$-year dataset, as performed to examine the evidence for a GWB. The digit in parentheses gives the uncertainty on the last quoted digit. All factors were computed with the Savage--Dickey formula [Eq.\ \eqref{eq:savagedickey}], with the hyperparameter priors listed in \autoref{tab:priors}.}
    \begin{tabular}{ld@{}dd@{}dd@{}dd@{}dd@{}d}
        \hline\hline
        & \multicolumn{2}{c}{uncorr.\ red process vs.\ pulsar noise} & \multicolumn{8}{c}{H.--D.\ corr.\ red process vs.\ pulsar noise} \\
\multicolumn{1}{l}{JPL ephemeris}
		& \multicolumn{2}{c}{(2A--1)} & \multicolumn{2}{c}{(3A--1)} & \multicolumn{2}{c}{with dipole (3B--2B)} & \multicolumn{2}{c}{with dipole, monopole (3C--2C)} & \multicolumn{2}{c}{with monopole (3D--2D)} \\        
        & \multicolumn{1}{c}{$\gamma=13/3$} & \multicolumn{1}{c}{$\gamma \in [0,7]$}
        & \multicolumn{1}{c}{$\gamma=13/3$} & \multicolumn{1}{c}{$\gamma \in [0,7]$}
        & \multicolumn{1}{c}{$\gamma=13/3$} & \multicolumn{1}{c}{$\gamma \in [0,7]$}
        & \multicolumn{1}{c}{$\gamma=13/3$} & \multicolumn{1}{c}{$\gamma \in [0,7]$}
        & \multicolumn{1}{c}{$\gamma=13/3$} & \multicolumn{1}{c}{$\gamma \in [0,7]$} \\ 
            \hline
            DE421           & 8.28(4)  & 5.3(2)  & 11.9(7)     & 6.5(2)  & 3.57(5)     & 2.07(6)     & 3.20(5)     & 1.96(5)   & 7.4(5)  &   3.7(3)     \\
            DE430           & 18.9(7)  & 8.7(4)  & 26\rlap{(2)}       & 12.8(9) & 3.69(4)     & 2.05(3)     & 3.94(9)     & 1.9(1)    & 12\rlap{(1)}   &   5.6(4)     \\
            DE435           & 1.82(4)  & 1.22(1) & 2.15(4)     & 1.69(5) & 1.52(2)     & 1.17(2)     & 1.35(2)     & 0.99(2)   & 1.77(4) &   1.43(4)    \\
            DE436           & 4.4(1)   & 3.5(2)  & 7.2(4)      & 4.8(2)  & 2.17(4)     & 1.54(3)     & 2.14(2)     & 1.34(4)   & 3.4(1)  &   2.18(5)    \\
            INPOP13c        & 24.9(7)  &  \mbox{---}   & \mbox{---}          & \mbox{---}      &  \mbox{---}        & \mbox{---}          &  \mbox{---}        &   \mbox{---}      & \mbox{---}      &   \mbox{---}         \\
        \hline
        \textsc{BayesEphem} & 0.884(9) & 0.647(7) & 1.00(2)    & 0.70(1) & 0.93(2)     & 0.67(3)     & 0.98(4)     & 0.66(2)   & 0.98(5)  & 0.70(3)      \\
            \hline
    \end{tabular}
\end{center}
\end{table*}%
\begin{table*}[ht]
	\begin{center}
	\scriptsize
	\caption{Bayes factors for model comparisons using NANOGrav's $11$-year dataset, as performed to examine the evidence for spatial correlations with different ORFs.	
	The digit in parentheses gives the uncertainty on the last quoted digit.
	All factors were computed with the product-space method discussed in Sec.\ \ref{sec:bayes}, with the hyperparameter priors listed in \autoref{tab:priors}.}
	\begin{tabular}{ld@{}dd@{}dd@{}d}
		\hline\hline
		& \multicolumn{2}{c}{3A--2A: H.-D.\ corr. red process}
		& \multicolumn{2}{c}{2B--2A: dipole corr. red process}
		& \multicolumn{2}{c}{2D--2A: monopole corr. red process} \\
JPL ephemeris		& \multicolumn{6}{c}{vs.\ uncorrelated red process} \\
        & \multicolumn{1}{c}{$\gamma=13/3$} & \multicolumn{1}{c}{$\gamma \in [0,7]$}
        & \multicolumn{1}{c}{$\gamma=13/3$} & \multicolumn{1}{c}{$\gamma \in [0,7]$}
        & \multicolumn{1}{c}{$\gamma=13/3$} & \multicolumn{1}{c}{$\gamma \in [0,7]$} \\
    		\hline
            DE421           & 1.34(7) & 1.53(8) & 0.46(3) & 0.60(3) & 0.18(1)  & 0.21(1)  \\
            DE430           & 1.44(8) & 1.7(1)  & 0.46(3) & 0.94(6) & 0.106(9) & 0.21(2)  \\
            DE435           & 1.24(6) & 1.42(7) & 0.55(3) & 0.85(4) & 0.54(3)  & 0.55(3)  \\
            DE436           & 1.45(8) & 1.63(9) & 0.57(3) & 1.05(6) & 0.46(3)  & 0.52(3)  \\
        \hline
        \textsc{BayesEphem} & 1.08(6) & 1.15(7) & 0.83(5) & 0.87(6) & 1.12(9)  & 0.96(7)  \\
    		\hline
	\end{tabular}
	\end{center}
	\label{tab:spatialcorr_bfs}
\end{table*}%

\begin{figure*}
  \includegraphics{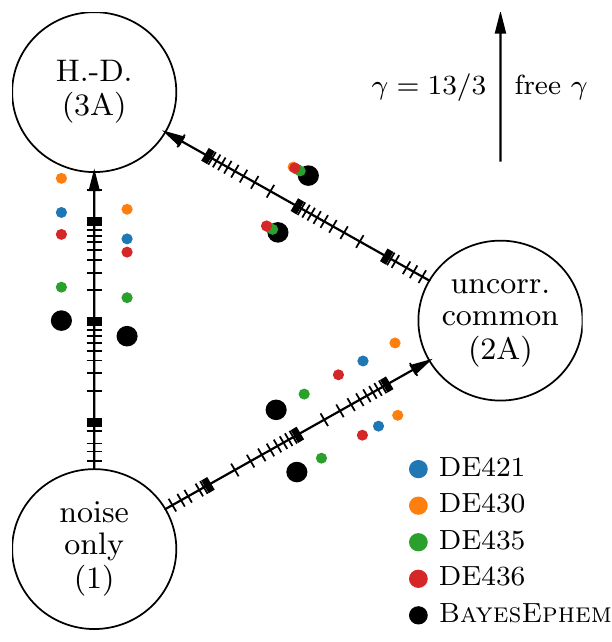}
  \includegraphics{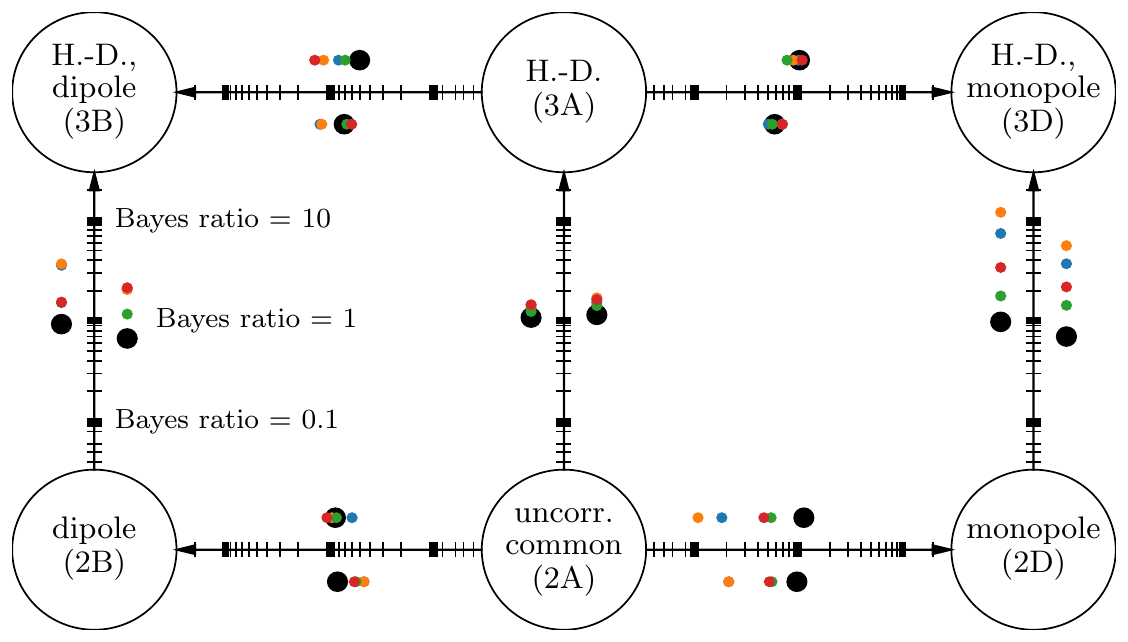}
  \caption{Bayes factors for model comparisons on the 11-year dataset: on the left, evidence of a GWB; on the right, effects of spatially correlated systematics. In these graphs, each model (as described in \autoref{tab:spatialcorr_modeltab}) is represented by a bubble, and for each pair of models the dots mark on a logarithmic scale the measured Bayes factor in favor of the model at the head of the arrow. Thus, dots are closer to the model favored by the data. The smaller colored dots represent Bayes factors computed by taking one of the DE421, DE430, DE435, and DE436 JPL ephemerides as a fixed-parameter model without uncertainties; the larger black dots represent Bayes factors computed by marginalizing over ephemeris errors (i.e., by adopting \bayesephem). Dots to the left of the arrows correspond to fixing the spectral slope $\gamma$ of the GWB to 13/3, as appropriate for a background from SMBHBs evolving purely by GW emission; dots to the right correspond to marginalizing over $\gamma$, taken to have uniform prior distribution in $[0,7]$. The graph on the left shows that when adopting the JPL ephemerides as fixed-parameter models, most of the evidence for a GWB accrues from the presence of unexplained red-spectrum residuals in each pulsar (2A--1), with a smaller preference added by modeling Hellings--Downs correlations (3A--2A); neither conclusion is supported by \bayesephem. As for the graph on the right: The bottom row compares a common uncorrelated red process with dipolar and monopolar processes; the former is favored. The top row examines the case for dipolar and monopolar processes in the presence of a Hellings--Downs (GW-like) signal. Comparing the vertical arrows in the left and right graphs we see that (for fixed JPL ephemerides) evidence for a GW-like signal is weakened when the model allows for other spatially correlated processes.}
  \label{fig:modelcomp}
\end{figure*}

In Tables \ref{tab:table_sd_bfs} and \ref{tab:spatialcorr_bfs} and in Fig.\ \ref{fig:modelcomp}, we show Bayes factors for two sets of model comparisons performed on the 11-year dataset to quantify the statistical evidence for a stochastic GWB and for coherent sources of systematic errors that lead to spatially correlated residuals.
The first four columns of \autoref{tab:table_sd_bfs} and the graph on the left of Fig.\ \ref{fig:modelcomp} are diagnostic of the multilevel decision scheme outlined above in Sec.\ \ref{sec:bayes}. Adopting the JPL ephemerides as fixed-parameter models, the data favor the presence of a common uncorrelated process in all pulsars, to various degrees and especially so for DE430, and they favor slightly the presence of Hellings--Downs inter-pulsar correlations. However, this preference disappears if we marginalize over the ephemeris uncertainties.

\begin{figure}
\includegraphics[width=\columnwidth]{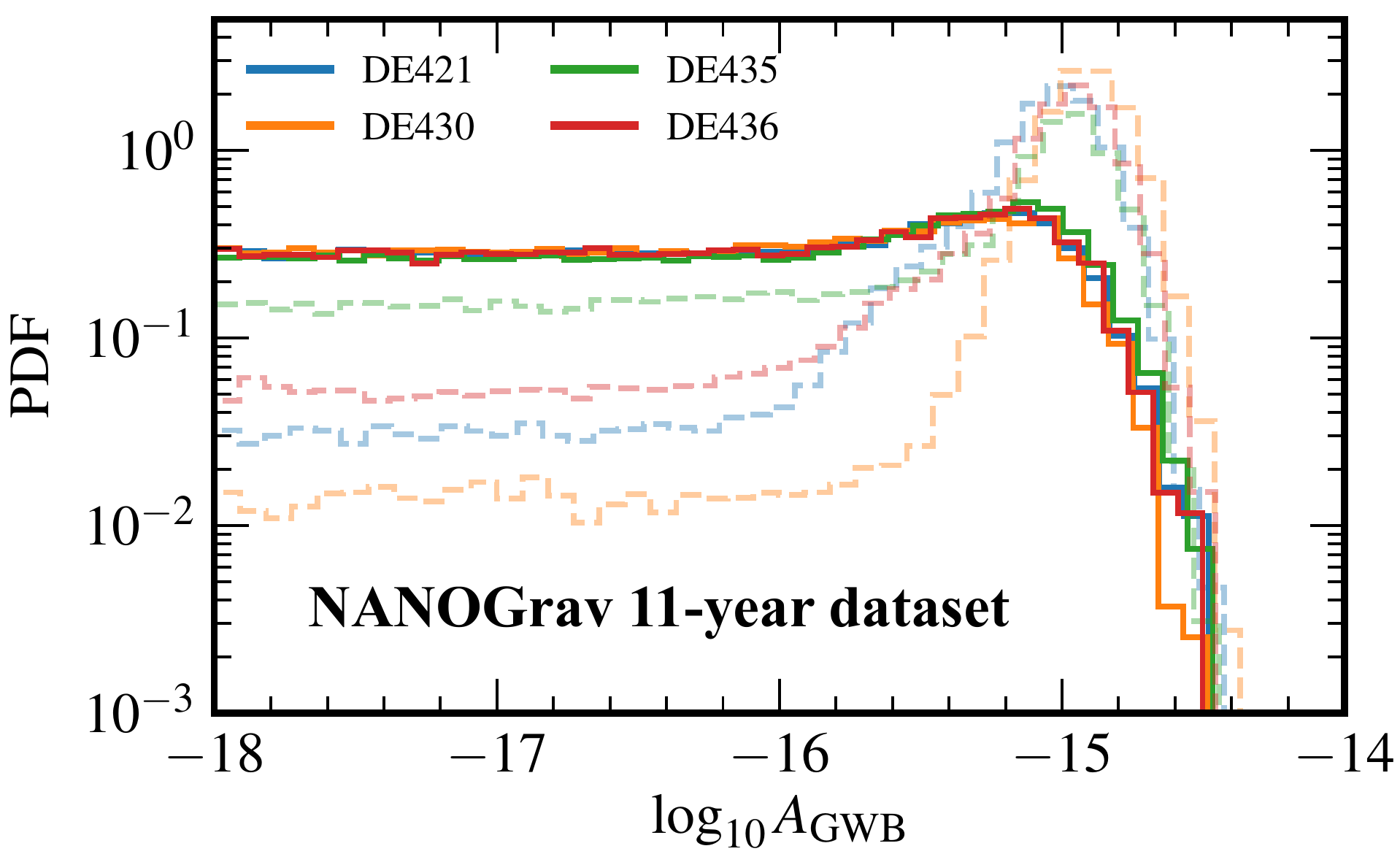}
  \caption{Posterior probability distributions for $A_\mathrm{GWB}$ (log-uniform prior, $\gamma = 13/3$, and no spatial correlations), as computed for the NANOGrav 11-year dataset under individual JPL ephemerides (dashed lines), and with \bayesephem, taking each of the JPL ephemerides as a starting point (solid lines). This plots demonstrates that \bayesephem\ bridges the JPL ephemerides successfully; in doing so it removes most evidence for the presence of a GWB.}
  \label{fig:nano11_ephconnect}
\end{figure}
The effects of ephemeris errors are also apparent in the upper plot of \autoref{fig:nano11_ephconnect}, which shows the posterior distribution of $\log_{10} A_\mathrm{GWB}$ under the log-uniform prior used to compute Bayes factors, for $\gamma = 13/3$, and neglecting Hellings--Downs correlations.
The dashed lines show the posterior obtained by taking each ephemeris as fixed-parameter models without uncertainties; the solid lines show the posteriors obtained by marginalizing over ephemeris uncertainties, starting with each ephemeris in turn. Although the dashed curves agree roughly in their modes, they have surprisingly different widths and \emph{contrast}, which we may define as the ratio of the peak posterior to its value in the lowest-amplitude (leftmost) bin; the latter appears in the denominator of the Savage--Dickey estimate [Eq.\ \eqref{eq:savagedickey}] of the signal-vs-noise Bayes factor.

The convergence of the solid lines to a flatter common shape demonstrates that our modeling of ephemeris uncertainties bridges the four ephemerides successfully, removing spurious evidence for GWs, or potentially absorbing a true GW signal. However, if a true GW signal is present, it happens to be significantly covariant with the systematic differences in the Roemer delays induced by the last few ephemerides; furthermore, the signal appears to weaken as we shift from older (DE421, DE430) to newer, plausibly more accurate ephemerides (DE435, DE436), although this trend is not entirely consistent. In this paper, we do not attempt to quantify whether these circumstances are realized often in the ensemble of possible datasets similar to ours; nevertheless, these circumstances motivate our choice of marginalizing over ephemeris uncertainties as the principled Bayesian strategy for our analysis.
 
The six rightmost columns of \autoref{tab:table_sd_bfs}, as well as \autoref{tab:spatialcorr_bfs} and the graph on the right of \autoref{fig:modelcomp}, document the degree to which the data favor the presence of timing-residual components with different spatial correlations. Components with both dipolar (ephemeris-error--like) and monopolar (clock-error--like) correlations are disfavored, although this conclusion is significantly weakened if we marginalize over ephemeris uncertainties. At the same time, the evidence for quadrupolar (GWB-like) correlations is weakened when the model allows for other spatially correlated processes. This is not unexpected, since spatial correlations with different multipolar structures only become truly orthogonal in the limit of many equally low-noise pulsars.

Indeed, discrimination of monopolar, dipolar, and quadrupolar correlation signatures will improve as our datasets gain more and more pairs of high--timing-precision pulsars with a broad distribution of angular separations. We plan to characterize discrimination requirements (on pulsar number, timing quality, and sky position) in our upcoming paper on SSE error modeling.

\paragraph{Impact of SSE error modeling on GW detection}
We performed a small number of simulations to test the impact of \textsc{BayesEphem} on our GWB detection prospects over the next few years.
To this end, we produced realistic $15$-yr datasets\footnote{To produce the datasets, we used actual observation epochs for the $34$ NANOGrav pulsars, and set residuals equal to white measurement noise plus red-spectrum intrinsic noise, at levels consistent with those estimated for the actual data (\citetalias{abb+17}).
We rescaled TOA uncertainties by a factor $1.5$, which calibrates the noise-only simulated dataset so that its $11.4$ year ``slice'' has the same (DE$436$, model $2$A) GWB upper limit as the real data.
We extended the dataset baseline to $15$ years by drawing observation epochs and TOA measurement errors from distributions of these quantities over the last $3$ years of real data.}
using DE$436$ and injecting GWBs of various amplitudes, and we analyzed the full datasets, as well as their $11.4$ yr ``slices,'' using DE$430$ and \bayesephem. We chose DE$430$ because it led to the highest signal-vs.-noise Bayes factor (model $2$A-vs.-$1$) and upper limits for the actual data.

\begin{figure}
\includegraphics[width=\columnwidth]{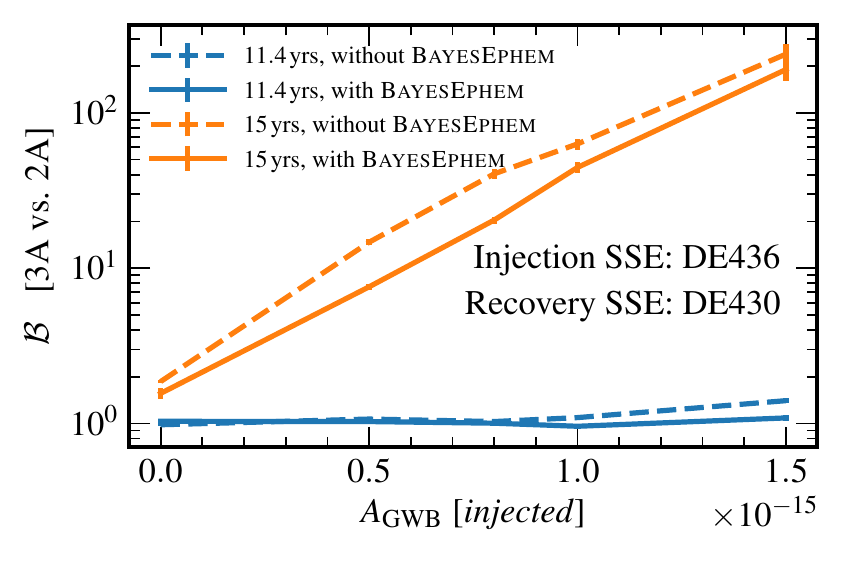}
\caption{Model $3$A-vs.-$2$A (spatial correlation) Bayes factors for a set of simulated $15$ yr datasets with GWB injections at different levels.  We analyze the full datasets (orange lines) as well as their $11.4$ yr ``slices'' (blue), both by adopting the ``wrong'' ephemeris (dashed) and by employing \textsc{BayesEphem} to marginalize over SSE errors (solid). We conclude that \textsc{BayesEphem} will not impede the ability of PTAs to make a definitive detection in the near future (see main text).}
  \label{fig:bayesephem_sims}
\end{figure}

For a noise-only simulation, we find that unmodeled systematic offsets between DE$436$ and DE$430$ are interpreted as a common red-spectrum process with a signal-vs.-noise Bayes factor (model $2$A-vs.-$1$) of $\sim 2$ in $11.4$ years of data, and $\sim 20$ in $15$ years of data. By contrast, \textsc{BayesEphem} is able to account for the offsets, reducing Bayes factors to levels consistent with noise fluctuations.
As we increase the injected GWB amplitude, model $2$A-vs.-$1$ Bayes factors remain low for $11.4$ yrs of data, even for amplitudes comparable to our fiducial upper limits.
The same is true for model $3$A-vs.-$2$A Bayes factors (the definitive spatial-correlation test for GWBs), which are plotted in \autoref{fig:bayesephem_sims}.

For $15$ years of data, the scaling of Bayes factors with injected GWB amplitude is comparable for both DE$430$ and \textsc{BayesEphem}. Remarkably, the potential covariance of \textsc{BayesEphem} parameters with GWB amplitude does not inhibit signal detection in the near future, even at astrophysically-pessimistic levels ($\sim 5\times 10^{-16}$, consistent with \citet{s16}).
Thus, while SSE errors may spuriously produce early signs of a GWB (i.e. a common red-spectrum process), their mitigation with \textsc{BayesEphem} will not impair prospects for near-future GWB detection. We regard our simulations as conservative, since additional pulsars, as well as improved timing precision and SSE accuracy, will accelerate progress toward detection.

\subsection{Optimal statistic}


\begin{table}
	\begin{center}
	\caption{Optimal statistic $\hat{A}_{\rm gw}^2$ and associated SNR for the NANOGrav 11-year dataset, assuming a $\gamma=13/3$ power-law GWB with Hellings--Downs spatial correlations. 
	The noise-marginalized computation provides a more accurate assessment of the significance of a 
	common red process compared to the fixed-noise due to the covariance between pulsar red-noise parameters and common red-noise parameters.
	(see Sec.\ \ref{sec:optimalstatdef}).}
	\label{tab:os_HD}
	\begin{tabular}{@{}lcccc@{}}
		\hline\hline
		&  \multicolumn{2}{c}{fixed noise} & \multicolumn{2}{c}{noise marginalized} \\
		JPL ephemeris		&  $\hat{A}_\mathrm{GWB}^2$ 	& SNR & mean $\hat{A}_\mathrm{GWB}^2$ & mean SNR		 \\
    		\hline
    		DE421 					& $\phantom{-}\num{8.23}{-31}$ 		& $\phantom{-}1.06\phantom{0}$ 		& $\num{8.9}{-31}$ 		& $0.9(9)$ 	\\
    		DE430 					& $\phantom{-}\num{2.32}{-31}$ 		& $\phantom{-}0.390$ 	& $\num{6.9}{-31}$ 		& $0.4(4)$	\\
    		DE435 					& $\num{-3.46}{-31}$ 	& $-0.640$ 	& $\num{5.9}{-31}$ 		& $0.7(6)$	\\
    		DE436 					& $\phantom{-}\num{4.47}{-32}$ 		& $-0.069$ 	& $\num{9.7}{-31}$ 		& $0.8(7)$	\\
    		\hline
	\end{tabular}
	\end{center}
\end{table}


\begin{table*}[ht]
	\begin{center}
	\caption{Noise-marginalized optimal statistic $\hat{A}_{\rm gw}^2$ and associated SNR for the NANOGrav 11-year dataset, assuming a $\gamma=13/3$ power-law GWB with Hellings--Downs (GW-like), monopolar (clock-error--like), and dipolar (ephemeris-error--like) spatial correlations. None of the SNRs are significant.}
	\label{tab:os_spatial}
	\begin{tabular}{@{}lcccccc@{}}
		\hline\hline
		& \multicolumn{2}{c}{Hellings--Downs} & \multicolumn{2}{c}{monopole} & \multicolumn{2}{c}{dipole} \\
		JPL ephemeris	& mean $\hat{A}_{\rm GWB}^2$ & mean SNR & mean $\hat{A}_{\rm GWB}^2$ & mean SNR & mean $\hat{A}_{\rm GWB}^2$ & mean SNR		 \\
    		\hline
    		DE421 					& $\num{8.9}{-31}$ 			& $0.9(9)$ & $\num{-6.2}{-33}$ 			& $0.0(6)$ & $\phantom{-}\num{3.8}{-32}$ 			& $0(1)$ 	\\
    		DE430 					& $\num{6.9}{-31}$			& $0.4(4)$	& $\phantom{-}\num{1.5}{-31}$ 			& $0.5(4)$& $\phantom{-}\num{2.4}{-31}$ 			& $0.7(9)$\\
    		DE435 					& $\num{5.9}{-31}$ 			& $0.7(6)$	& $\phantom{-}\num{8.5}{-32}$ 			& $0.5(5)$& $\phantom{-}\num{5.7}{-32}$ 			& $1(1)$\\
    		DE436 					& $\num{9.7}{-31}$ 			& $0.8(7)$	& $\phantom{-}\num{2.0}{-31}$ 			& $0.9(7)$& $\phantom{-}\num{1.9}{-31}$ 			& $1(1)$\\
    		\hline 
		\textsc{BayesEphem}		& $1.3\times 10^{-31}$ 				& $0.1(9)$ 		&  $\phantom{-}2.7\times 10^{-32}$	& $0(1)$ & $-4.3\times 10^{-32}$ & $0(1)$ \\
		\hline
	\end{tabular}
	\end{center}
	\label{default}
\end{table*}

\autoref{tab:os_HD} compares the fixed-noise and noise-marginalized optimal statistic (see Sec.\ \ref{sec:optimalstatdef}) for a Hellings--Downs spatially correlated common process 
computed using DE421, DE430, DE435, and DE436. 
The noise-marginalization was performed using 10\,000 realizations of the noise. 
Except for DE421, the fixed-noise analysis systematically underestimates $\hat{A}_\mathrm{GWB}^2$ and SNR compared to the noise-marginalized analysis 
because of the covariance between pulsar red-noise parameters and the common red-noise parameters.
Note that, although the optimal statistic is formulated in terms of the squared amplitude, negative $\hat{A}_\mathrm{GWB}^2$ and SNR values are possible if noise fluctuations result in negative correlations. 
In the noise-marginalized analysis, we find mean SNR $< 1$ for all ephemerides---no appreciable evidence of Hellings--Downs correlations. These results are consistent with the Bayesian analysis. 

\autoref{tab:os_spatial} compares the noise-marginalized optimal statistic computed for Hellings--Downs spatial correlations with variants of the statistic that model dipolar and monopolar correlations. 
In addition to computing the optimal statistic using individual ephemerides, we also 
use \bayesephem\ to marginalize over the ephemeris uncertainty. 
For all of these analyses, we find no evidence for a common process with either Hellings--Downs, monopolar, or dipolar spatial correlations.

\begin{figure}
  \includegraphics[width=\columnwidth]{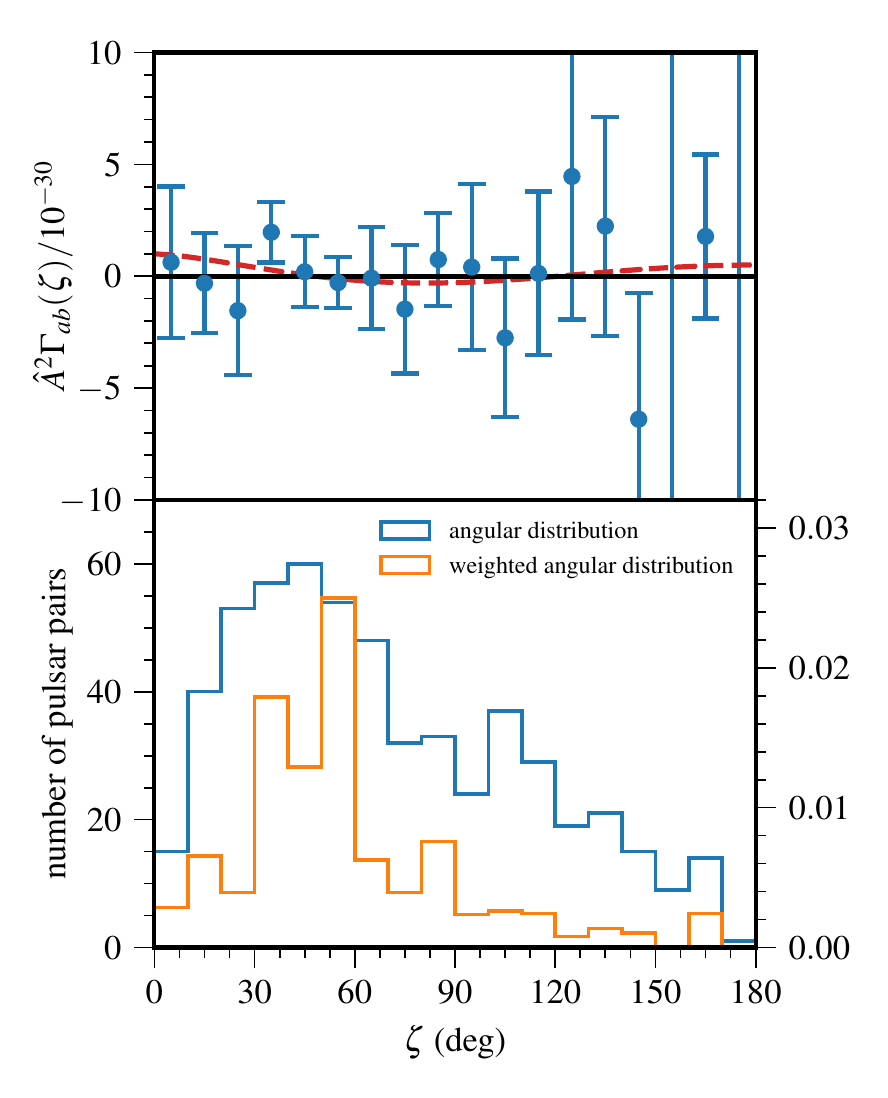}
  \caption{\textit{Top:} Angular distribution of cross-correlated power between pulsar pairs in the NANOGrav 11-year dataset, averaged over 10-degree bins. This analysis was done using DE436; other ephemerides give similar results. 
 A GWB would cause the cross-correlated power to lie along the Hellings--Downs curve (red dashed line), shown assuming a GWB amplitude of $A_\mathrm{GWB} = 10^{-15}$. 
  		\textit{Bottom:} Histogram of pulsar-pair angular separations. The blue curve shows numbers in each bin, while the orange curve is reweighted by squared 1-$\sigma$ uncertainties of the averaged cross-correlated power in that bin. Currently NANOGrav is most sensitive at angular separations between $30^\circ$ and $60^\circ$.
}
  \label{fig:os_angulardist}
\end{figure}
The upper half of \autoref{fig:os_angulardist} shows the mean noise-marginalized cross-correlated power between pulsar pairs as a function of angular distribution, averaged into 10 degree bins.
There is no evidence of the Hellings--Downs correlations characteristic of isotropic GWBs.
The lower half of the plot shows a histogram of angular separations for the pulsar pairs in our dataset: NANOGrav is currently most sensitive to angular separations between $30^\circ$ and $60^\circ$, which correspond to the smallest errors in the cross-correlation plot.

\subsection{Comparison of 9-year and 11-year results}
\label{sec:9to11}

\begin{figure}
\includegraphics[width=\columnwidth]{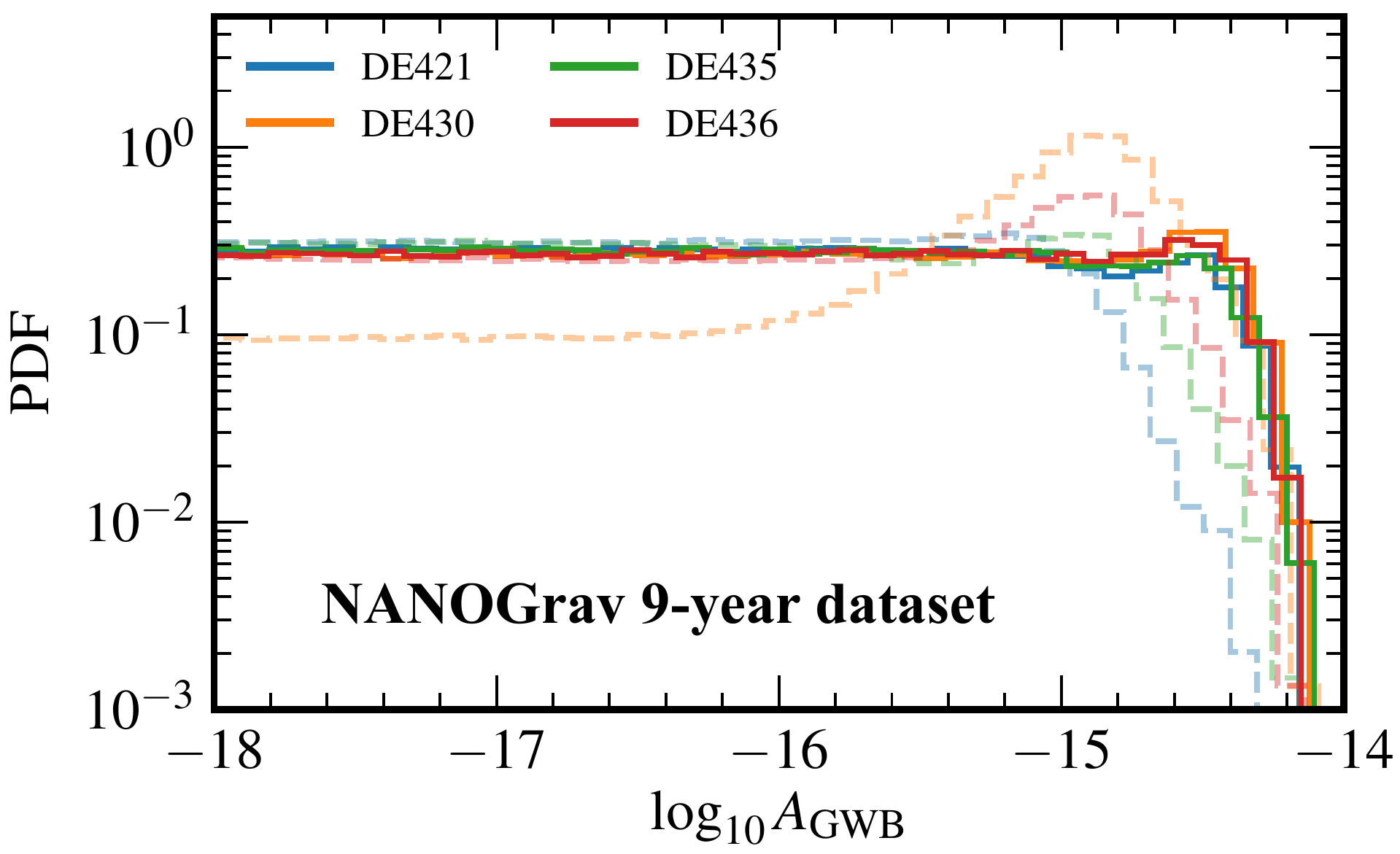}
  \caption{Posterior probability distributions for $A_\mathrm{GWB}$ (log-uniform prior, $\gamma = 13/3$, and no spatial correlations), as computed for the NANOGrav 9-year dataset under individual JPL ephemerides (dashed lines), and with \bayesephem, taking each of the JPL ephemerides as a starting point (solid lines).}
  \label{fig:nano9_ephconnect}
\end{figure}
The 9-year analysis of \citetalias{abb+16} adopted DE421 as a fixed-parameter model without uncertainties, and did not include Hellings--Downs correlations. Thus, a straight comparison can be made with the 11-year DE421 model-2A results: the $\gamma = 13/3$ upper limit remains at $1.5 \times 10^{-15}$, while the $\gamma = 13/3$ Bayes factor vs.\ pulsar noise changes from $0.81$ to $8.3$; however, this comparison is not very significant given what we have learned about ephemeris errors.

Applying \bayesephem\ to the 9-year dataset successfully bridges $A_\mathrm{GWB}$ posteriors (see \autoref{fig:nano9_ephconnect}), and yields a model-2A (uncorrelated) upper limit of $2.67(2) \times 10^{-15}$ and a model-3A (H.--D.-correlated) upper limit of $2.91(2) \times 10^{-15}$ (both for $\gamma = 13/3$). Thus, our fiducial model-3A upper limit improves by a factor $2.91/1.45 = 2.0$ in the 11-year dataset. This is greater than expected from simple scaling arguments \citep{sejr13}, for which the additional two years of data should reduce the limit from $2.91\times 10^{-15}$ to $1.85\times 10^{-15}$, i.e.\ an improvement of $\sim 1.6$. The major cause of this discrepancy is presumably that the longer 11-year baseline is better able to disentangle ephemeris perturbations, which have typical timescales of the $11.86$-year Jupiter period. 

The model-2A Bayes factors vs.\ pulsar noise are $0.910(7)$ for $\gamma = 13/3$ and $1.210(4)$ for $\gamma\in [0,7]$, while they are $1.27(1)$ and $2.29(3)$ for model 3A. All Bayes factors under \bayesephem\ are comparably uninformative for the 9-year and 11-year datasets.

\begin{figure}
 \centering
 \includegraphics[width=0.5\textwidth]{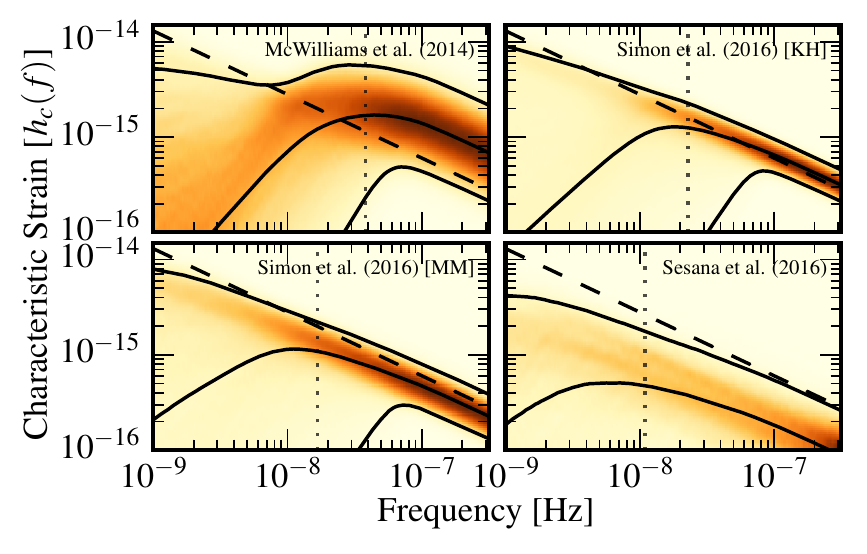}
  \caption{Posterior density plots of broken--power-law GWB spectra, as constrained by the $11$-year dataset, adopting high-frequency $A_\mathrm{GWB}$ priors from \citetalias{mop14}, \citet{ss16}, \citetalias{s16}. At each frequency, a thin vertical segment of the plot should be understood as a density plot of the characteristic strain; the solid lines mark the $2.5\%$, $50\%$, and $97.5\%$ quantiles, the dashed line shows the $\gamma = 13/3$ unbroken--power-law upper limit, and the vertical dotted lines show the median $f_{\rm bend}$ values. All plots were produced for a spatially uncorrelated common GWB process. Larger values of $A_\mathrm{GWB}$ induce stronger turnovers at higher frequencies, (e.g. \citetalias{mop14} has a median value [$50\%$ quantile] of $f_\mathrm{bend}$ that is more than three times that of \citet{s16}). }
  \label{fig:bayesograms}
\end{figure}
We also reproduce the spectral-turnover analysis of \citetalias{abb+16}, which models the GWB with a broken--power-law spectrum [Eq.\ \eqref{eq:broken}, following \citet{scm15}]. \autoref{fig:bayesograms}, obtained with $A_\mathrm{GWB}$ priors from \citetalias{mop14}, \citet{ss16}, \citetalias{s16} (using \bayesephem\ and neglecting H.--D.\ correlations) can be contrasted with Fig.\ 5 of \citetalias{abb+16}. In both figures, models that infer larger GWB levels require turnovers at higher GW frequencies to be consistent with the data: the \citetalias{mop14} prior gives a median value of $f_\mathrm{bend}$ at $3.83 \times 10^{-8}$ Hz, while the \citetalias{s16} prior gives a value of $1.09 \times 10^{-8}$ Hz, a difference of more than a factor of three. This analysis is a useful tool in broadly understanding various models' consistency with PTA limits. However, it is limited by attributing a single value of $\kappa$ [see \autoref{eq:broken}] to the entire population of SMBHBs, and it is unable to incorporate eccentricity, which flattens out the turn-over and skews towards higher GW frequencies.  
In Sec.\ \ref{sec:discussion} we present results incorporating a more sophisticated approach \citep{tss17} allowing us to confront astrophysical population models directly.

\section{Limits on Astrophysical Models}
\label{sec:discussion}

Some of the most exciting science made possible by the NANOGrav data is realized when we use the GWB constraints to confront the astrophysics of various source populations.
Now, the most likely source population for PTAs is SMBHBs.
In \citetalias{abb+16} we introduced simple PTA constraints on SMBHB population parameters, but due to methodological limitations we were unable to deliver a realistic analysis, i.e.\ we derived constraints for the parameters describing a broken--power-law spectrum, and then reinterpreted those constraints in terms of SMBHB effects that could alter the spectrum, taken one at a time.
In this paper we adopt the modeling framework developed by \citet{tss17} to go much further: we use a set of population-synthesis simulations to explore the effects of population parameters on the GWB spectrum, then constrain those population parameters directly from the data.
We apply the same method also to the most recent cosmic-string models.

\subsection{Supermassive black-hole binaries}
\label{sec:smbhb}

PTAs are sensitive to the stochastic GWB comprised of the superposition of GWs from merging SMBHBs throughout the Universe. The details of this background (i.e., spectral shape and amplitude) are sensitive to the physics of SMBHB evolution. The history of SMBHB mergers is generally assumed to follow the history of galaxy mergers, but the exact relation remains an open question.
Dynamical friction initially causes SMBHs to sink toward each other in a post-merger galaxy remnant, but becomes an inefficient means of further hardening at parsec separations \citep{bbr80}. Additional dynamical influences are required to drive a SMBHB to milliparsec orbital separations, and thus into the PTA frequency band. This supposed \textit{final parsec problem} can be overcome by a variety of processes: $(i)$ three-body scattering effects with stars in the galaxy's bulge, where stars in the binary's loss cone slingshot off the binary carrying away orbital energy \citep{mv92,q96,shm06}; $(ii)$ interactions between the binary and a viscous circumbinary disk \citep{ipp99,hkm09,ks11}; $(iii)$ eccentricity, which increases the rate of binary evolution \citep{pm63,p64} and can be amplified by $(i)$ \citep{shm06,s10,rs11} and $(ii)$ \citep{an05,c+09,rds+11}.
If the \textit{final parsec problem} is not completely overcome by additional environmental processes, a subsequent galaxy merger may add a third massive black hole to the system, which can drive the initial binary towards coalescence and may increase that binary's eccentricity \citep{bsb+17, rph+18}.
All of these influences can cause the shape of the GWB spectrum in the PTA band to deviate from the fiducial $f^{-2/3}$ power-law at low frequencies ($f\in[1,10] ~\mathrm{nHz}$) causing a change in slope or a turnover if the binary remains coupled to the environment or has large orbital eccentricities \citep{enn07,ks11,s13c,rws+14,scm15,hmgt15,tss17}. 

As discussed earlier, we use a Gaussian-process spectral model to explore the parameter space of SMBHB environments and dynamics. We perform sophisticated population-synthesis simulations over a $5\times 5\times 5$ grid in the $\{\alpha_\mathrm{BH}, \rho_\mathrm{stars}, e_0\}$ parameter space, where $\alpha_\mathrm{BH}$ is the $y$-intercept of the $M_\mathrm{BH}-M_\mathrm{bulge}$ relationship, $\rho_\mathrm{stars}$ is the typical mass density of galactic-core stars at the binary influence radius, and $e_0$ is binary eccentricity at formation.
At each grid point we perform $100$ simulations and compute the mean spectrum and uncertainty from Poisson variation.
We then train a GP at each GW frequency, allowing spectral amplitudes to be predicted with uncertainties over the entire parameter space. These predictions act as priors on the strain within the free-spectrum model. We set uniform priors on the \textit{astrophysical parameters} corresponding to $\alpha_\mathrm{BH}\in\{7,9\}$, $\log_{10}[\rho_\mathrm{stars}/M_\odot\mathrm{pc}^{-3}]\in\{1,4\}$, and $e_0\in\{0,0.95\}$.

\begin{figure}
 \centering
 \includegraphics[width=0.45\textwidth]{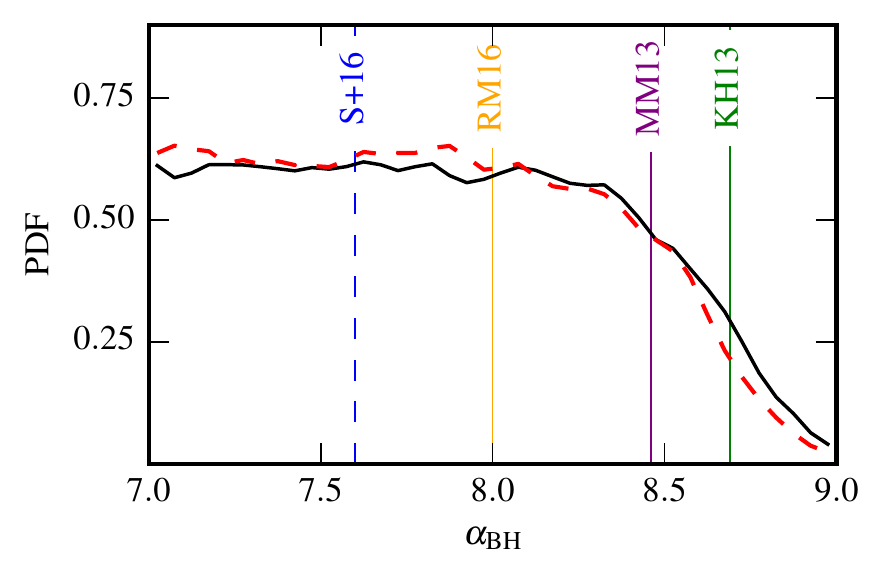}
  \caption{Constraints on $\alpha_\mathrm{BH}$ from the NANOGrav 11-year dataset. The black solid line is the posterior PDF marginalized over the combined parameter space $\{\rho_\mathrm{stars}, e_0 \}$, while the red dashed line is the posterior PDF marginalized over $\rho_\mathrm{stars}$ for circular binaries $(e_0 = 0)$. The red line is slightly more constraining, which is to be expected when a degree of freedom is removed. We do not set a $95\%$ upper limit from these posteriors, since that number would be dependent on the lower bound of the $\alpha_\mathrm{BH}$ prior. The colored lines show selected observational measurements and predictions for $\alpha_\mathrm{BH}$: KH13 \citep{kh13}, MM13 \citep{mm13}, RM17 \citep{rm17b}, S+16 \citep{sbs+16}. The S+16 line is dashed because that measurement is not a simple power-law relation, but includes higher-order terms; here we base our plot on the leading-order coefficient.}
  \label{fig:alpha_lim}
\end{figure}

\begin{figure*}
    \centering
    \subfloat[][2D posterior for $\rho_\mathrm{stars}$ and $e_0$ for fixed $\alpha_\mathrm{BH}$]{\includegraphics[width=\textwidth]{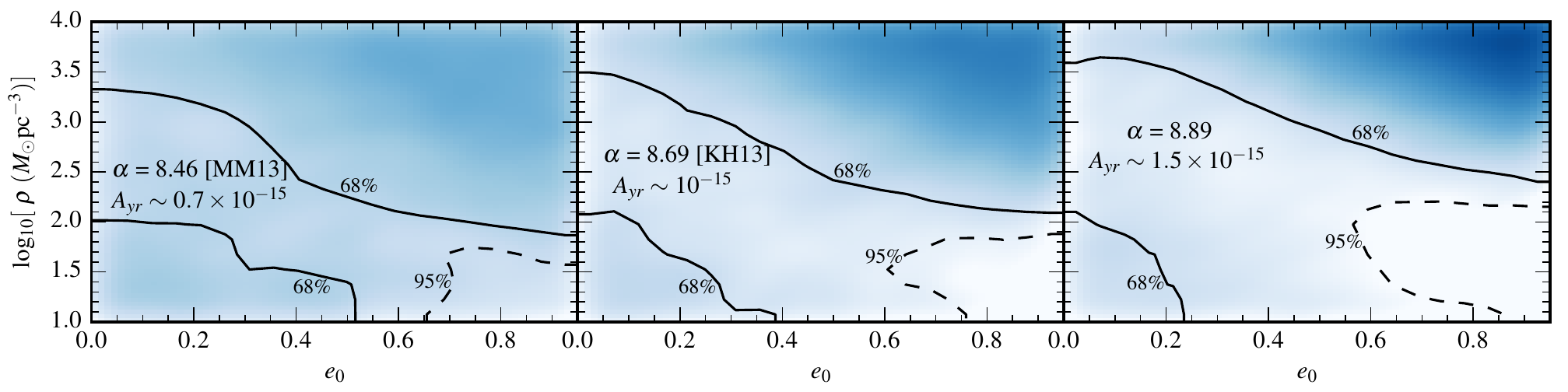}
   	\label{fig:GP_rhoecc}}
    \\
    \subfloat[][Marginalized spectral shape of GWB]{
        \includegraphics[width=\textwidth]{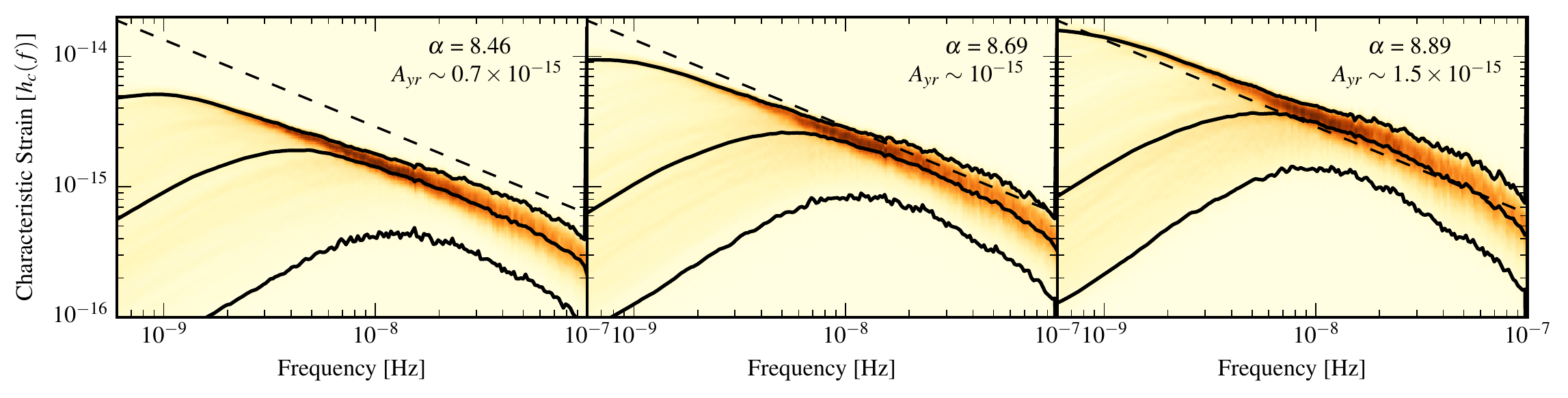}
        \label{fig:GP_bayes}} 
    \caption{\emph{Top (a)}: posteriors for $\rho_\mathrm{stars}$ and  $e_0$ at different values of $\alpha_\mathrm{BH}$, as computed for the NANOGrav 11-year dataset. \emph{Bottom (b)}: marginalized spectral densities computed from those posteriors. Each column of plots corresponds to a different value of $\alpha_\mathrm{BH}$ with values decreasing from right to left. The center and left columns correspond to the measured values from \citet{kh13} and \citet{mm13}, respectively, while the right column corresponds to a larger value, for comparison. The solid (dashed) line in (a) corresponds to the $68\% ~(95\%)$ contour and the blue shading is consistent across all of the plots. The dashed and solid lines in (b) are identical to those shown in \autoref{fig:bayesograms}, where the dashed line indicates our upper limit on $A_\mathrm{GWB}$ of $1.34(1) \times 10^{-15}$ on a power law GWB [$f^{-2/3}$] and the solid lines show the $2.5\%$, $50\%$, and $97.5\%$ confidence levels. As $\alpha_\mathrm{BH}$ increases, so to does the overall level of the background, and the spectral shape of the GWB is more constrained by the data. 
    }
    \label{fig:rho_vs_ecc}
\end{figure*}

Our population-synthesis model is similar to the scheme described in \citet{tss17,ss16}, where the SMBHB merger rate density was constructed from observed galaxy properties and SMBH--host-galaxy relations. Specifically, we adopt a galaxy stellar mass function from \citet{iml+13}, a galaxy pairing fraction from \citet{rdd+14}, and a parametrized $M_\mathrm{BH}-M_\mathrm{bulge}$ relationship. The $M_\mathrm{BH}-M_\mathrm{bulge}$ relation is set by three parameters: $\alpha_\mathrm{BH}$, $\beta_\mathrm{BH}$ and $\epsilon_\mathrm{BH}$, where $\log_{10}(M_\mathrm{BH} / M_\odot) = \alpha_\mathrm{BH} + \beta_\mathrm{BH}\log_{10}(M_\mathrm{bulge} / 10^{11} M_\odot)$ and $\epsilon_\mathrm{BH}$ is the intrinsic scatter of points around the set power law. We fix $\beta_\mathrm{BH} = 1$ and $\epsilon_\mathrm{BH} = 0.3$, values typical of observational measurements \citep[see e.g.][]{kh13,mm13}. As shown in \citet{ss16}, $\alpha_\mathrm{BH}$ is the parameter of maximal impact on the GWB; as such, it is the only parameter from the $M_\mathrm{BH}-M_\mathrm{bulge}$ relation that is varied in this work. However, there are impacts on the level of the GWB predicted from changing all of these parameters, which are explored in depth in \citet{ss16}, and therefore the limits on $\alpha_\mathrm{BH}$ in this work must be interpreted in that context.

The eccentricity evolution in this model follows the prescription first derived in \citet{q96}, and later expanded upon in \citet{s10}. However, recent work in \citet{rm17a} \citep[see also][]{sgd11,gds12,mtk+17} has shown that eccentricity evolution can be damped by the rotation of the central stellar bulge, which would lessen the effect of extreme initial eccentricities.

The parameter $\alpha_\mathrm{BH}$ primarily changes the overall level of the GWB, while $\rho_\mathrm{stars}$ and $e_0$ primarily change its spectral shape. We start to explore this parameter space by constraining $\alpha_\mathrm{BH}$. \autoref{fig:alpha_lim} shows $\alpha_\mathrm{BH}$ posteriors derived by  marginalizing over $\{\rho_\mathrm{stars}, e_0 \}$ (black solid line), and by marginalizing over $\rho_\mathrm{stars}$ for circular sources $(e_0 = 0)$ (red dashed line). The constraint for circular sources is slightly more stringent, as is expected from removing a degree of freedom. However, in both cases the determination of $\alpha_\mathrm{BH}$ in \citet[][hereafter \citetalias{kh13}]{kh13} is disfavored compared to the others.

Quantitatively, we may take the ratios of PDFs as a proxies for model-comparison Bayes factors between $\alpha_\mathrm{BH}$ determinations: by doing so, we find \citet[][hereafter \citetalias{mm13}]{mm13} to be $1.5$ times more probable than \citetalias{kh13}, while the other two measurements are $1.9$ times more probable than \citetalias{kh13}. These constraints become $2$ and $2.6$, respectively, for circular sources $(e_0 = 0)$. As stated above, these results for $\alpha_\mathrm{BH}$ need to be viewed in the context of the complete model used to infer the population of SMBHBs,
which relies on the assumption that the SMBH merger rate follows the observed galaxy merger rate. This may not be the case if the \textit{final parsec problem} is not solved for all systems, or if binary evolution takes much longer than anticipated by this model \citep{tgv+18}.
However, even when we robustly incorporate many of the parameters that impact the spectral shape of the GWB, the NANOGrav 11yr dataset prefers values of $\alpha_\mathrm{BH}$ that are lower than the largest observed measurements from \citetalias{kh13}.

We can also compute a joint marginalized posterior for $\rho_\mathrm{stars}$ and $e_0$, but the $\alpha_\mathrm{BH}$ distribution is too broad for this to be useful.  It is more informative to examine $p(\rho_\mathrm{stars},e_0|\alpha_\mathrm{BH})$ for a few representative values of $\alpha_\mathrm{BH}$. In \autoref{fig:rho_vs_ecc}, we show $\alpha_\mathrm{BH} = 8.46$, $8.69$, and $8.89$. The first two values are the measurements reported in \citetalias{mm13} and \citetalias{kh13}, while the third is an even larger value.
The top panels of \autoref{fig:rho_vs_ecc} show posteriors, while the bottom panels show the corresponding marginalized spectral distributions, using the same conventions as \autoref{fig:bayesograms}.
As $\alpha_\mathrm{BH}$ increases from left to right, the GWB increases in level, and its spectral shape needs to deviate more strongly from a $f^{-2/3}$ power law to be consistent with the data. This trend is seen also in the increased preference towards larger $\rho_\mathrm{stars}$ and $e_0$. While this effect was observed in earlier work, the methodology used in this paper allows for its robust exploration.

Taken at a glance, the results detailed in this work appear less constraining then those presented in \citetalias{abb+16}. This is to be expected: marginalizing over parameters, rather than fixing them to set values, will insert more uncertainty into any constraint. Additionally, the methods used in \citetalias{abb+16} incorporated an intermediate step by extrapolating from the posterior on $A_\mathrm{GWB}$ while assuming a power-law GWB. In this paper, we are able to constrain the entire spectrum directly from the dataset with no intermediaries---a benefit of the GP method of \citet{tss17}, which will enable future NANOGrav datasets to place constraints on the dynamics of the most massive black holes in the Universe.

\subsection{Cosmic strings}

Cosmic strings are linear topological defects that can form in the early Universe as a result of symmetry-breaking phase transitions \citep{k76,v81,v85,Vilenkin:2000jqa}.
Strings that form with lengths greater than the horizon are known as ``long'' or ``infinite'' strings, while smaller strings form loops.
If two strings meet one another they can exchange partners, and small portions of string can be chopped off with a \textit{reconnection} probability $p$.
For classical strings $p=1$, but String-Theory--inspired models may have $p<1$. This is due to the fact that fundamental strings interact probabilistically, and also that in these models an intersection occurring in the usual three spatial dimensions need not occur in higher compactified dimensions. Cosmic string networks evolve toward an attractor solution known as the ``scaling regime'' in which the statistical properties of the system (such as the average size of loops or the distance between long strings) scale with the cosmic time, and the energy density of the string network is a small constant fraction of the radiation or matter density. Cosmic strings have tensions equal to their mass per unit length, $\mu$.
This tension is so high that strings oscillate relativistically under their own tension, decaying solely through the emission of GWs, and shrinking in size.
The formation of loops and their subsequent decay by GW emission is the mechanism by which the string network loses energy and reaches the scaling regime. The GW spectrum from cosmic string networks is exceptionally broadband, covering all regions of LIGO, LISA, and PTA sensitivity.
For our purposes, we describe the parameter space of cosmic strings in terms of their dimensionless tension, $G\mu/c^2$, and their reconnection probability, $p$.

We take a more self-consistent approach than previous PTA analyses.
Rather than re-fit posterior samples (from power-law or free spectrum searches) to cosmic string models \citep{abb+16,ltm+15}, we train a GP interpolant on output from the most up-to-date string population simulations.
\citet{bo17} and \citet{bos17} performed a complete end-to-end calculation of the stochastic GW background expected from a network of cosmic strings, namely: $(i)$ simulation of the long-string network to find a representative sample of loop sizes and shapes; $(ii)$ modeling of loop shape deformations due to gravitational back-reaction; $(iii)$ GW spectrum computed for each loop; $(iv)$ evaporation and production modeled to find the distribution of loops over $z$; $(v)$ integration of the GW spectrum of each loop over the redshift-dependent loop distribution; and finally $(vi)$ integration over cosmological time to find the present-day GW background.

The output from these simulations corresponds to GW energy density spectra at a range of string tension values, $G\mu/c^2$, over $25$ orders of magnitude in frequency and has been made publicly available.\footnote{\href{http://cosmos.phy.tufts.edu/cosmic-string-spectra/}{http://cosmos.phy.tufts.edu/cosmic-string-spectra/}} We convert these to characteristic strain, then at each frequency-bin in our PTA analysis we train a GP to emulate the strain as a function of string tension.
We expand our model to include reconnection probability, $p$, by analytically scaling the fiducial $p=1$ strain spectrum by $(1/p)^{1/2}$ \citep{sak05}.
We then use this model (with all features of the cosmic-string spectrum included) to analyze the NANOGrav 11-year dataset. We do not model signal finiteness or anisotropy due to bright resolvable cosmic-string bursts, since this is only expected when initial loop sizes are very small ($\lesssim 10^{-8}$) \citep{ktyk17}.

\begin{figure}
 \centering
 \includegraphics[width=0.5\textwidth]{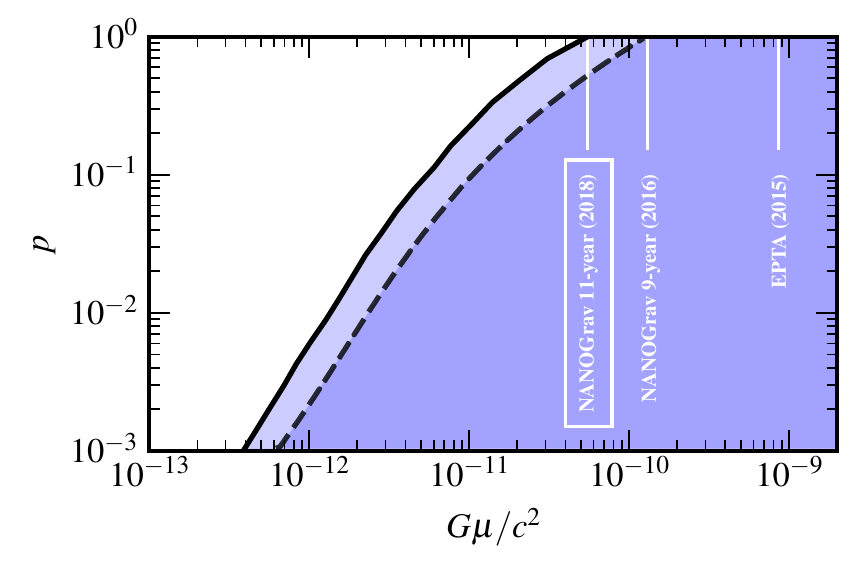}
  \caption{Constraints on cosmic-string tension, $G\mu/c^2$, as a function of reconnection probability, $p$, with the NANOGrav 11-year dataset.
    The excluded region of parameter space is bounded by a solid black-line.
    The corresponding excluded region for the NANOGrav 9-year dataset \citepalias{abb+16} is bounded by a dashed black line, while the EPTA constraints \citep{ltm+15} are shown for $p=1$ only.}
  \label{fig:cosmicstrings}
\end{figure}

\autoref{fig:cosmicstrings} shows the $95\%$ upper limit on string tension as a function of reconnection probability.
The shaded region enclosed by the solid black line indicates parameter space that is excluded by the NANOGrav 11-year dataset under the assumptions of the \citet{bo17} cosmic string simulations.
For $p=1$ the string tension is constrained to be $G\mu/c^2< 5.3(2)\times 10^{-11}$. At this level we would not expect any measurable effects in the CMB power spectrum, nor through gravitational lensing \citep{bos17}. PTAs are currently the best experiment with which to detect cosmic strings, and to place stringent limits on the string parameter space.

By contrast, the NANOGrav 9-year dataset \citepalias{abb+16} constraints on string tension (shown as an excluded region with a dashed line boundary) were computed under the assumptions of older string simulations \citep{bjo+14}, and were obtained by re-sampling the posterior distribution of a power-law GWB spectrum.
For $p=1$ the string tension was constrained to be $G\mu/c^2< 1.3\times 10^{-10}$.
Finally, even though the most recent EPTA constraints on cosmic strings 
\citep{ltm+15} were not computed under the assumptions of the \citet{bjo+14} simulations, in \citetalias{abb+16} the constraints were converted to get a corresponding limit on the string tension of $G\mu/c^2 < 8.6\times 10^{-10}$.
Thus, the constraints on cosmic string tension from the NANOGrav 11-year dataset are $2.5$ times better than the NANOGrav 9-year dataset, and $16.2$ times better than the most recent EPTA analysis. The $9$- to $11$-year improvement is to be expected, since \bayesephem\ analyses of the $11$-year dataset give consistently more constraining GWB limits than DE$421$ analyses of the $9$-year dataset.
There are a few other notable caveats to these comparisons; $(i)$ the NANOGrav $9$-year and EPTA analyses were performed under a fixed JPL SSE model, while the NANOGrav $11$-year analysis uses \bayesephem; $(ii)$ the simulation advances of \citet{bo17} with respect to \citet{bjo+14} impede a direct comparison.
However, the additional $\sim 2$ years of data in the new NANOGrav dataset, the new SSE uncertainty modeling, and the improved end-to-end analysis with simulated cosmic-string spectra all combine to increase NANOGrav's sensitivity to the cosmic-string parameter space.

\subsection{Primordial gravitational-waves}
\label{sec:primordial}

According to the theory of inflation, quantum fluctuations in the spacetime geometry of the early Universe are amplified to cosmological scales. Inflation leaves a background of relic primordial GWs that may be observable today \citep{g76, g77, s80, l82, fp83}.
Studies of the cosmic microwave background (CMB) that attempt to observe these GWs indirectly through their imprint of tensor-mode CMB polarizations are limited to probing the surface of last scattering, roughly 300,000 years after the Big Bang \citep{kks97, sz97, bpa+15}.
By contrast, GW observations can in principle observe a much earlier epoch in the history of the Universe, extending back to as little as $10^{-32}$ s post Big Bang.
Indeed, the spectral index of the primordial GWB is determined by the equation-of-state parameter $w$ in the immediate post-inflation, pre--Big-Bang-Nucleosynthesis Universe, and by the tensor index $n_t$, which depends on the detailed dynamics of inflation (see \citet{g05} and references therein).
The primordial spectral dependencies are typically stated in terms of GWB-$\alpha$ as in \citet{lms+16} and \citetalias{abb+16}.
We can express GWB-$\gamma$ (see Eq.\ \ref{eq:specdef}) for a primordial spectrum as
\begin{equation}
    \gamma = \frac{4}{3w+1} + 3 - n_t.
\end{equation}

In \autoref{tab:primordialUL} we list Bayesian 95\% upper limits on $A_\mathrm{GWB}$ of a primordial GWB, derived as described in Sec.\ \ref{sec:bayes}.
We consider three scenarios, the same considered in \citetalias{abb+16}, fixing $\gamma$ to values corresponding to each:
radiation-dominated ($w=1/3$), matter-dominated ($w=0$), and kinetic-energy--dominated ($w=1$) equations of state.
Following \citet{z11}, we assume a scale-invariant primordial power spectrum (i.e., $n_t=0$) for all cases.

These limits constrain the energy density spectrum of the primordial GWB by way of
\begin{equation} \label{eq:omega_gwb}
    \Omega_\mathrm{GWB}(f)\, h^2 = \frac{2\pi^2}{3 {H_0}^2} f^2 \, {h_c}^2(f),
\end{equation}
where $h$ is the dimensionless Hubble parameter, $H_0=100$~km$\,$s$^{-1}$$\,$Mpc$^{-1}$, and $h_c$ is the characteristic GW strain.
For a radiation-dominated post-inflationary Universe, we obtain
\begin{equation} \label{eq:omega_flat}
    \Omega_\mathrm{GWB}(f_\mathrm{yr})\,h^2 \le 3.4(1) \times 10^{-10},
\end{equation}
after marginalizing over SSE uncertainties.
This is a 20\% improvement over the result quoted in \citetalias{abb+16}; that number, however, should be revised upward significantly due to SSE bias. Referring back to the bottom panel of \autoref{fig:hcf}, we see that the energy-density sensitivity of our PTA dataset is dominated by the lowest few frequencies, which individually have $95\%$ upper limit values of $\sim 10^{-9}$, but which in combination beat the limit down to the value quoted in \autoref{eq:omega_flat}.


\begin{table}[t]
	\scriptsize
	\caption{NANOGrav 11-year upper limits on primordial GWs (last digit uncertainty): 95\% credible intervals obtained under uniform priors for GW and pulsar--red-noise amplitudes, and quoted at reference frequency $f_\mathrm{yr}=\mathrm{yr}^{-1}$.}
	\begin{center}
	\begin{tabular}{@{} cccccc @{}}
		\hline\hline
    		\multirow{2}{*}{Ephemeris}	& \multicolumn{3}{c}{$95\%$ upper limit on $A_\mathrm{GWB}$ [$\times 10^{-15}$]} \\
						& KE dom.\ $(\gamma=4)$ & Rad.\ dom.\ $(\gamma=5)$ & Mat.\ dom.\ $(\gamma=7)$ \\					
    		\hline
    		DE421 				& $2.01(3)$  & $0.81(1)$  & $0.100(3)$ \\
    		DE430 				& $2.32(2)$  & $0.92(1)$  & $0.117(2)$ \\
    		DE435 				& $2.04(2)$  & $0.84(1)$  & $0.105(2)$ \\
    		DE436 				& $2.10(2)$  & $0.88(1)$  & $0.111(1)$ \\
    		\hline
            \textsc{BayesEphem}		& $1.78(2)$  & $0.74(1)$  & $0.099(2)$ \\
    		\hline
	\end{tabular}
	\end{center}
	\label{tab:primordialUL}
\end{table}

\section{Summary and Conclusions}
\label{sec:conclusions}

This paper reports on the search for an isotropic stochastic GW background (GWB) in NANOGrav's $11$-year dataset. We targeted a GW signal with predominantly low-frequency power, and so analyzed only those pulsars that have greater than $3$ years of observations, corresponding to $34$ out of the $45$ in the data release. Our investigations encompassed different models of the GWB strain spectrum, spatial correlations between pulsars, and Solar System ephemeris (SSE). The latter influence was rigorously studied, and led to the major discovery of this paper: 

\begin{itemize}
\item We found significant variations in GW upper-limits and detection statistics when the dataset was analyzed under different published models of the SSE. These models are primarily from the Jet Propulsion Laboratory (JPL), ranging from DE$421$ to DE$436$. We also performed a limited analysis with INPOP$13$c.
\item For a model with Hellings--Downs spatial correlations between pulsars (as appropriate for an isotropic GW background), the $95\%$ upper limit on the amplitude of a fiducial $f^{-2/3}$ power-law strain spectrum (from an astrophysical population of SMBHBs) at a frequency of $1\,\mathrm{yr}^{-1}$ varies between $1.53$--$1.79\times 10^{-15}$. 
\item The ratio of Bayesian evidences between models that include a GWB versus only intrinsic pulsar noise processes varies between $\sim 2$ and $\sim 26$ in favor of a GWB, while the odds favoring GW-induced spatial correlation between pulsars vary between $1.18:1$ and $1.63:1$. The frequentist analog to the Bayesian odds-ratio (known as the ``optimal-statistic'') gives a signal-to-noise ratio for GW-induced spatial correlations that varies between $0.57$ and $0.87$.
\end{itemize}

This discovery has major ramifications on how we interpret previous PTA results, and also how our analysis methodology must be revised for future searches. 

\begin{itemize}
\item We formulated a perturbative model (``\textsc{BayesEphem}'') that acts to bridge the systematic offsets in the various published models of the SSE, resulting in the first pulsar-timing constraints on GWs that are robust against Solar System uncertainties. This model corrects for coordinate-frame drift, uncertainties in gas-giant masses, and uncertainties in Jupiter's orbital elements.
\item Under this new model, the upper limit on the strain amplitude becomes $1.34\times 10^{-15}$ for a common red-spectrum process, and $1.45\times 10^{-15}$ for a GWB. Adding further spatially-correlated processes in the model served to worsen these limits only slightly. 
\item The evidence ratio for models that include a GWB versus only intrinsic pulsar noise processes is $1$ for a GWB with fixed spectral slope, and $0.70$ if the spectral slope is varied. The odds ratio favoring GW-induced spatial correlations between pulsars is $1.08:1$ if the spectral slope is fixed, or $1.15:1$ if the slope is varied. The frequentist optimal-statistic gives a signal-to-noise ratio for GW-induced spatial correlations of $0.09$, where the spectral slope is necessarily fixed at the fiducial value of $-2/3$. Both the Bayesian and frequentist analysis show inconclusive evidence for a GW-like red-spectrum process and quadrupolar inter-pulsar spatial correlations.
\end{itemize}

We also performed a systematic study of spatially-correlated processes in the PTA dataset under different ephemerides, tabulating upper limits and evidence ratios for various combinations of a common red-spectrum process, GWB, stochastic clock error, and stochastic SSE uncertainty. With \textsc{BayesEphem} the presence of these additional spatially-correlated processes slightly worsens the GW upper limits, but all remain broadly consistent within uncertainties. Dipole spatial correlations between pulsars seem most disfavored under \textsc{BayesEphem}, likely because we have dealt with the most plausible source of such correlations with our deterministic SSE-uncertainty modeling. Uncertainties in the evidence and odds ratios (in addition to their absolute values being around unity) prevent us being able to make strong statements. The NANOGrav $11$-year dataset is only weakly informative of spatial correlations between pulsars.

We used the NANOGrav $11$-year dataset (with the \textsc{BayesEphem} model) to place constraints on the parameter space of astrophysical and cosmological sources of GWs. As in \citetalias{abb+16}, we placed priors on the high-frequency strain amplitude that are motivated by different SMBHB modeling scenarios, then allowed the presence of a turnover in the shape of the strain spectrum to be constrained by the data. With a positive GWB detection, signs of a spectral turnover could indicate that dynamical evolution of SMBHBs remains strongly driven by galactic environmental processes even at centiparsec orbital separations, thereby offering a solution to the ``final-parsec problem''. For a non-detection (as we currently have) this procedure also acts as a test of the validity of our high-frequency strain priors, i.e. priors with larger strain amplitude at high-frequency are more in tension with the data when extrapolated back to low frequencies via the fiducial $f^{-2/3}$ scaling, necessitating low-frequency spectral attenuation to ensure consistency with non-detection. We found that the \citetalias{mop14} prior led to a turnover within the sensitivity band of our PTA ($f > 1/T \sim 3\mathrm{nHz}$) with greater than $97.5\%$ credibility. Other astrophysically-motivated priors gave greater consistency with a pure power-law strain spectrum.

For this paper, we took a large step forward in GW spectral modeling and analysis. As described in \citet{tss17} we trained a Gaussian process (GP) model on strain spectra from SMBHB population simulations carried out over a large grid in astrophysical parameter space, namely the $y$-intercept of the $M_\mathrm{BH}-M_\mathrm{bulge}$ relation [$\alpha_\mathrm{BH}$], the typical mass density of stars in a galactic core ($\rho_\mathrm{stars}$), and the binary eccentricity at formation ($e_0$). This trained model acts as a prior on the GW strain at each frequency, allowing direct recovery of the posterior distribution of astrophysical parameters. We found that the NANOGrav 11yr dataset prefers values of $\alpha_\mathrm{BH}$ that are lower than the largest observed measurements from \citetalias{kh13}. Taking the ratios of probability densities as a proxy for model-comparison Bayes factors, we found \citetalias{mm13} to be $1.5$ times more probable than \citetalias{kh13}, while other, lower measurements are $1.9$ times more probable than \citetalias{kh13}. These constraints become $2$ and $2.6$, respectively, when we consider only circular sources (binary eccentricity at formation equaling zero). By studying different values of $\alpha_\mathrm{BH}$, we showed how larger levels of the GWB, which require spectral shapes that deviate more from the common power law, set progressively tighter constraints on the joint parameter space of $\{\rho_\mathrm{stars}, e_0 \}$. The modeling utilized to produce these results can be trivially expanded to incorporate new astrophysical complexity, and in the era of precision spectral characterization it will allow PTAs to construct a detailed view of SMBH demographics out to $z\sim 2$.

We took a similar modeling approach for strain spectra resulting from decaying cosmic string networks, where we calibrated a GP model with the simulations of \citet{bo17}. This gave an SSE-marginalized $95\%$ upper limit on the string tension of $G\mu/c^2=5.3\times 10^{-11}$ at a reconnection probability of $p=1$, which is $2.5$ times better than \citetalias{abb+16}, and $16.2$ times better than \citet{ltm+15}. (Although these previous published limits were computed without SSE uncertainty modeling). 
PTAs have already surpassed conventional cosmological probes of cosmic string networks \citep{lms+16}, and will continue to offer the best constraints for the foreseeable future. Likewise, we obtained a limit on a background of primordial GWs resulting from the inflation of quantum spacetime fluctuations (with a radiation-dominated post-inflationary Universe), corresponding to $\Omega_\mathrm{GWB}h^2 < 3.4\times 10^{-10}$ at $95\%$ credibility with SSE marginalization. This is a $20\%$ improvement over \citetalias{abb+16}, and an even larger improvement once proper SSE modeling is taken into account for the $9$-year analysis.

Over the last few years, the PTA community has made great strides in gathering ever larger, higher-quality datasets, and in developing sophisticated analysis methods that can deal with the complex noise budgets and subtle systematics typical of pulsar timing, while interfacing ever more closely and robustly with the astrophysics of GW sources. The sequence of recent stochastic-GW papers [for NANOGrav, \citet{dfg+13}, \citet{abb+16}, this paper] is a fitting witness to this growth. We expect this effort to be rewarded by nanohertz GW detection within the next several years \citep{2016ApJ...819L...6T}, if the steadfast pursuit of methodological rigor and physical insight remains our cynosure.

\acknowledgements

\emph{Author contributions.}
An alphabetical-order author list was used for this paper in recognition of the fact that a large, decade timescale project such as NANOGrav is necessarily the result of the work of many people.
All authors contributed to the activities of the NANOGrav collaboration leading to the work presented here, and reviewed the manuscript, text, and figures prior to the paper's submission.
Additional specific contributions to this paper are as follows.
ZA, KC, PBD, MED, TD, JAE, ECF, RDF, EF, PAG, GJ, MLJ, MTL, LL, DRL, RSL, MAM, CN, DJN, TTP, SMR, PSR, RS, IHS, KS, JKS, and WZ developed the eleven-year
data set through a combination of observations, arrival time calculations, data checks and refinements, and timing model development an analysis;
additional specific contributions to the data set are summarized in \citetalias{abb+17}.
SRT coordinated the writing of the paper and led the search.
SRT, JAE, PTB, KPI, SJV, TTP, JSH, and NSP directly ran the analysis pipelines.
SRT and MV designed the \textsc{BayesEphem} statistical analysis.
SJV, KPI, JAE designed and ran the optimal-statistic analysis, and interpreted the results.
EH participated in the optimization of (some) of the gravitational wave detection pipelines used in this analysis.
NJC, XS, MV, TJWL provided feedback on searches and new analysis techniques, as well as vetted the paper in an internal review process.
JS, SRT, and MV developed the interpretation of astrophysical results.
PTB ran the relic GW analysis and interpreted the results.
SRT, JS, and XS developed and interpreted the cosmic strings results.
SRT, JAE, JS, MV, PTB, TTP, JSH, MV, XS, SJV, KPI, CMFM wrote the paper, collected the bibliography, prepared figures and tables.

\emph{Acknowledgments.}
We thank the referee for useful suggestions and comments that improved the quality of this manuscript. The NANOGrav project receives support from NSF Physics Frontier Center award number 1430284.
NANOGrav research at UBC is supported by an NSERC Discovery Grant and Discovery Accelerator Supplement and by the Canadian Institute for Advanced Research. We thank our colleagues in the International Pulsar Timing Array for comments and useful discussions. We thank Alberto Sesana for commenting our astrophysical modeling and interpretation. 
MV and JS acknowledge support from the JPL RTD program.
Portions of this research were carried out at the Jet Propulsion Laboratory, California Institute of Technology, under a contract with the National Aeronautics and Space Administration.
SRT was partially supported by an appointment to the NASA Postdoctoral Program at the Jet Propulsion Laboratory, administered by Oak Ridge Associated Universities through a contract with NASA. SRT thanks ERS for fruitful discussions.
JAE was partially supported by NASA through Einstein Fellowship grants PF4-150120.
SBS was supported by NSF award \#1458952.
PTB acknowledges support from the West Virginia University Center for Gravitational Waves and Cosmology. MAM was partially supported by NSF award OIA-1458952. WWZ is supported by the CAS Pioneer Hundred Talents Program and the Strategic Priority Research Program of the Chinese Academy of SciencesGrant No. XDB23000000. RvH was supported by NASA Einstein Fellowship grant PF3-140116.
This work was supported in part by National Science Foundation Grant No.~PHYS-1066293 and by the hospitality of the Aspen Center for Physics.
Portions of this work performed at NRL are supported by the Chief of Naval Research.
This research was performed in part using the Zwicky computer cluster at Caltech supported by NSF under MRI-R2 award No.~PHY-0960291 and by the Sherman Fairchild Foundation.
A majority of the computational work was performed on the Nemo cluster at UWM supported by NSF grant No.~0923409.
Parts of the analysis in this work were carried out on the Nimrod cluster made available by S.M.R.
Data for this project were collected using the facilities of the National Radio Astronomy Observatory and the Arecibo Observatory.
The National Radio Astronomy Observatory is a facility of the NSF operated under cooperative agreement by Associated Universities, Inc. The Green Bank Observatory is a facility of the National Science Foundation operated under cooperative agreement by Associated Universities, Inc.
The Arecibo Observatory is operated by SRI International under a cooperative agreement with the NSF (AST-1100968), and in alliance with Ana G.~M\'endez-Universidad Metropolitana and the Universities Space Research Association.
This research is part of the Blue Waters sustained-petascale computing project, which is supported by the National Science Foundation (awards OCI-0725070 and ACI-1238993) and the state of Illinois.
Blue Waters is a joint effort of the University of Illinois at Urbana-Champaign and its National Center for Supercomputing Applications.
Some of the algorithms used in this article were optimized using the Blue Waters allocation ``Accelerating the detection of gravitational waves with GPUs''.
The Flatiron Institute is supported by the Simons Foundation.

\bibliographystyle{yahapj}
\bibliography{apjjabb,bib}

\end{document}